\documentclass[aps,prb,reprint,superscriptaddress]{revtex4-2}
\usepackage{amsfonts}
\usepackage{amsmath}
\usepackage{amssymb}
\usepackage{graphicx}
\usepackage{epstopdf}
\usepackage{color}
\usepackage{bbold}
\usepackage[colorlinks, citecolor=red]{hyperref}

\renewcommand{\vec}[1]{\boldsymbol{#1}}

\begin{document}

\title{Open quantum theory of magnetoresistance in mesoscopic magnetic materials}

\author{Xian-Peng Zhang}

\affiliation{International Center for Quantum Materials, Beijing Institute of Technology, Zhuhai, 519000, China}

\affiliation{Centre for Quantum Physics, Key Laboratory of Advanced Optoelectronic Quantum Architecture and Measurement (MOE), School of Physics, Beijing Institute of Technology, Beijing, 100081, China}

\affiliation{Department of Physics, Hong Kong University of Science and Technology, Clear Water Bay, Hong Kong, China}

\author{Xiangrong Wang}

\affiliation{Department of Physics, Hong Kong University of Science and Technology, Clear Water Bay, Hong Kong, China}

\author{Yugui Yao}
\email{ygyao@bit.edu.cn}
\affiliation{Centre for Quantum Physics, Key Laboratory of Advanced Optoelectronic Quantum Architecture and Measurement (MOE), School of Physics, Beijing Institute of Technology, Beijing, 100081, China}

\affiliation{International Center for Quantum Materials, Beijing Institute of Technology, Zhuhai, 519000, China}

\begin{abstract}
Magnetoresistance (MR) in magnetic materials arises from spin-exchange coupling between local moments and itinerant electrons, representing a challenging many-body open-quantum problem. Here we develop a comprehensive microscopic theory of MR within an open-quantum-system framework by solving the Liouville–von Neumann equation for a hybrid system of free electrons and local moments using the time-convolutionless projection operator method. Our approach reveals both ferromagnetic and antiferromagnetic MR as consequences of temperature- and field-dependent spin decoherence, encompassing spin relaxation and dephasing. In particular, the resistance associated with spin decoherence is governed by  the order parameters of magnetic materials, such as the magnetization in ferromagnets and the Néel vector in antiferromagnets. This theory deepens the fundamental understanding of MR and offers guidance for interpreting and designing experiments on magnetic materials.
\end{abstract}

\maketitle

\section{Introduction}

Magnetoresistance (MR)—the change of a material’s electrical resistance under an applied magnetic field—provides a powerful probe of spin-dependent scattering and underpins key spintronic functionalities from magnetic sensing to nonvolatile memory. Fascinating effects, including anisotropic MR~\cite{van2002temperature,ramos2008anomalous,miao2024anisotropic,rushforth2007anisotropic,zeng2020intrinsic,dai2022fourfold,tu2025orbital}, Hanle MR~\cite{dyakonov2007magnetoresistance,velez2016hanle}, magnon MR~\cite{khvalkovskiy2009high,nguyen2011detection}, giant MR~\cite{binasch1989enhanced,baibich1988giant}, and tunnel MR~\cite{julliere1975tunneling,miyazaki1995giant,moodera1995large}, give valuable insights into the magnetic properties of various materials and offer tremendous potential for
applications in spintronics technology~\cite{chappert2007emergence}. The MR effect, which stems from the spin-exchange coupling between local moments and itinerant electrons in magnetic materials, is a challenging many-body and open-quantum problem. 

The conventional microscopic theories of MR in ferromagnets primarily focused on the plasma frequency and momentum relaxation time of itinerant electrons~\cite{ritzinger2023anisotropic}. For example, the 
anisotropic MR ($\rho _{\text{AMR}}\cos^2\alpha$), i.e., the change of resistance with the orientation ($\alpha$) of the magnetization relative to the electric current direction~\cite{mcguire1975anisotropic,campbell1970the,ebert1996anisotropic,smit1951magnetoresistance,van1959anisotropy,trushin2009anisotropic,wang2023theory,wang2024theory}, originates from i) the anisotropic plasma frequency caused by the interplay of magnetization and spin-orbit coupling~\cite{velev2005ballistic,kato2008intrinsic,vyborny2009semiclassical,nadvornik2021Broadband} and ii) anisotropic momentum relaxation time arising from either the magnetization-dependent scattering rate of itinerant electrons~\cite{mcguire1975anisotropic,campbell1970the,ebert1996anisotropic,smit1951magnetoresistance,van1959anisotropy,trushin2009anisotropic} or the magnon population of the local moments~\cite{goodings1963electrical,yosida1957anomalous,raquet2002electron,mihai2008electron} (i.e., magnon MR~\cite{khvalkovskiy2009high,nguyen2011detection}). 
Additionally, the Hanle spin precession of the spin accumulation leads to Hanle MR~\cite{dyakonov2007magnetoresistance,velez2016hanle}. 
Even though these theories have established the foundations for understanding Hanle and anisotropic MR, they can not quantitatively, or even qualitatively, analyze the diverse MR effects observed in experiments especially involving complicated dependencies
on the strength of the magnetic field ($\vec{B}$) and temperature ($T$)~\cite{van2002temperature,ramos2008anomalous,miao2024anisotropic,rushforth2007anisotropic,hupfauer2015emergence,lin2014experimental,nguyen2011detection,xiao2015four,miao2021magnetocrystalline,bason2009magnetoresistance}. For instance, i) the previous Hanle MR theory excludes the $B$- and $T$-dependent spin-exchange field [i.e., Eq.~\eqref{elf}] that contributes to the Hanle spin precession, and therefore is incapable of adequately including $B$ and $T$ modulation of Hanle MR; ii) the previous  anisotropic MR theory is mainly based on purely phenomenological formula, that is, $\rho _{\text{AMR}}$ is phenomenological parameters independent of $B$ and $T$. Here, we unveil that previous theories have overlooked a crucial aspect - the \textit{strong} $B$ and $T$ dependencies of the itinerant electrons' spin precession [i.e., Eq.~\eqref{elf}] and relaxation [i.e., Eqs.~(\ref{LongiSRT}) and (\ref{TransSRT})] that are ubiquitous in magnetic materials~\cite{zhang2019theory,gomez2020strong,oyanagi2021paramagnetic} and thus must play a crucial role in MR.

In this work, we develop a comprehensive microscopic theory of MR from an open-quantum system perspective. Specially, we solve the Liouville-von Neumann equation of the hybrid system consisting of free electrons and local moments through time-convolutionless projection operator technique. Both ferromagnetic and antiferromagnetic magnetoresistance are presented through the temperature- and field-dependent spin decoherence (i.e., spin relaxation and spin dephasing), where resistances change with the magnetization and the Néel vector, respectively. Our theory contributes to a deeper understanding of the fundamental physics underlying MR and provides insights for experiments involving magnetic materials.

The paper is organized as follows. In Sec.~\ref{systemHamiltonian},  we show the Hamiltonian of the hybrid system of free electrons and local moments.  In Sec.~\ref{openquantumtheory}, based on open quantum system framework, we solve the Liouville–von Neumann equation using the time-convolutionless projection operator method, where we attain anisotropic spin relaxation, a key ingredient of our MR effects. Sec.~\ref{diffusionequation} presents our anisotropic spin diffusion equation with boundary conditions caused by the spin Hall effect.  Sec.~\ref{resultsanddiscussions} are our results and discussions, 
including ferromagnetic MR  (Sec.~\ref{ferromagneticMR}) and antiferromagnetic MR (Sec.~\ref{antiferromagneticMR}). Our paper ends with conclusion. The derivations of many-body collision integral and ferromagnetic resistance are provided in Appendix~\ref{fvadkfvk} and Appendix~\ref{resistivity}, respectively. Finally, we also study the MR effect of normal metal decorated with magnetic impurities in Appendix~\ref{abgvkfk}.

\section{System Hamiltonian} \label{systemHamiltonian}
The MR effect, at its core, deals with the interplay between local moments and itinerant electrons. We  consider a ferromagnetic material described by the many-body Hamiltonian 
\begin{align} \label{dafvgsb}
    H=H_{e}+H_{m}+V_{em}.
\end{align}
The itinerant electrons' Hamiltonian  is given by 
\begin{align}
    H_e&=p_l^2/2m+\textsl{g}\mu _{B}B \hat{\vec{b}}\cdot\vec{\mathbf{\sigma}}_l\\
    &-(\hbar/4m^2c^2)\vec{\sigma}_l \cdot [\vec{p}_l \times \vec{\nabla} V_{\text{so}}(\vec{r}_l)]\notag,
\end{align}
with  
$\omega _{B}=\textsl{g}\mu _{B}B/\hbar$, where $\hat{\vec{b}}=\vec{B}/B$ is the unit vector of the  uniform external
field $\vec{B}$, $\textsl{g}\simeq 2$ is the $g$-factor, $\mu _{B}$ is the Bohr magneton, $\hbar$ is the reduced Planck constant, $\vec{\sigma}_l=(\sigma_l^x,\sigma_l^y,\sigma_l^z)$ is the vector of Pauli matrices representing the $n$th itinerant electron with position $\vec{r}_i$ and momentum $\vec{p}_i$, and $V_{\text{so}}(\vec{r}_i)$ includes both intrinsic and extrinsic spin–orbit potential~\cite{winkler2003spin}. Hereafter,  repeated indices are summed over. We also include a term describing the coupling of  the local moments themselves and with external magnetic field 
\begin{align} \label{bvabggb}
   \widehat{H}_{m}=g\mu _{B}\vec{B}\cdot \sum_{\imath i}\widehat{\vec{S}}_{\imath i}- \sum_{\imath i,\imath' i'}J_{\imath i,\imath' i'}\widehat{\vec{S}}_{\imath i}\cdot \widehat{\vec{S}}_{\imath' i'}, 
\end{align}
where  $\vec{S}_{\imath i}$ is a spin operator situated on site $i$ of sublattice $\imath$, $g$ is the $g$-factor,
 and $J_{\imath i,\imath' i'}=J(\vec{r}_{\imath i}-\vec{r}_{\imath' i'})$ is the
exchange energy between spins on sites $\vec{r}_{\imath i}=\vec{R}_{i}+\vec{r}_{\imath}$
and $\vec{r}_{\imath' i'}=\vec{R}_{i'}+\vec{r}_{\imath'}$.
It is assumed that all of the atoms on sublattice $\imath$ are identical.  
Finally, $V_{em}$ describes the effect of the local moments with scattering potential~\cite{mott1936electrical,mott1936resistance,mott1964electrons,goodings1963electrical,yosida1957anomalous,raquet2002electron,mihai2008electron,zhang2023extrinsic,zhang2022microscopic}
\begin{equation} \label{spin-exchange}
\check{V}_{\mathrm{em}}=-\hbar\sum_{\vec{i}l\nu}\widehat{L}^{\nu}_{\vec{i}}\hat{V}^{\nu}_{\vec{i}l}, 
\end{equation}
with 
\begin{equation}  \label{ElectronPart}
 \hat{V}^{\nu}_{\vec{i}l}=\hat{\sigma}^{\nu}_l\delta(\vec{r}_l-\vec{r}_{\vec{i}}),
\end{equation}
\begin{align}  \label{llectronPart}
    \widehat{L}^{\nu}_{\vec{i}}=\frac{\mathcal{J}^{\nu}_{\imath}}{2\hbar}  \widehat{S}_{\imath i}^{\nu},
\end{align}
where $\vec{i}=(\imath,i)$, $\nu=x,y,z$, and $\mathcal{J}^{\nu}_{\imath}$ describes the anisotropic  spin exchange coupling strength between itinerant electrons and local moments.

\begin{figure}[t!]
\begin{center}
\includegraphics[width=0.43\textwidth]{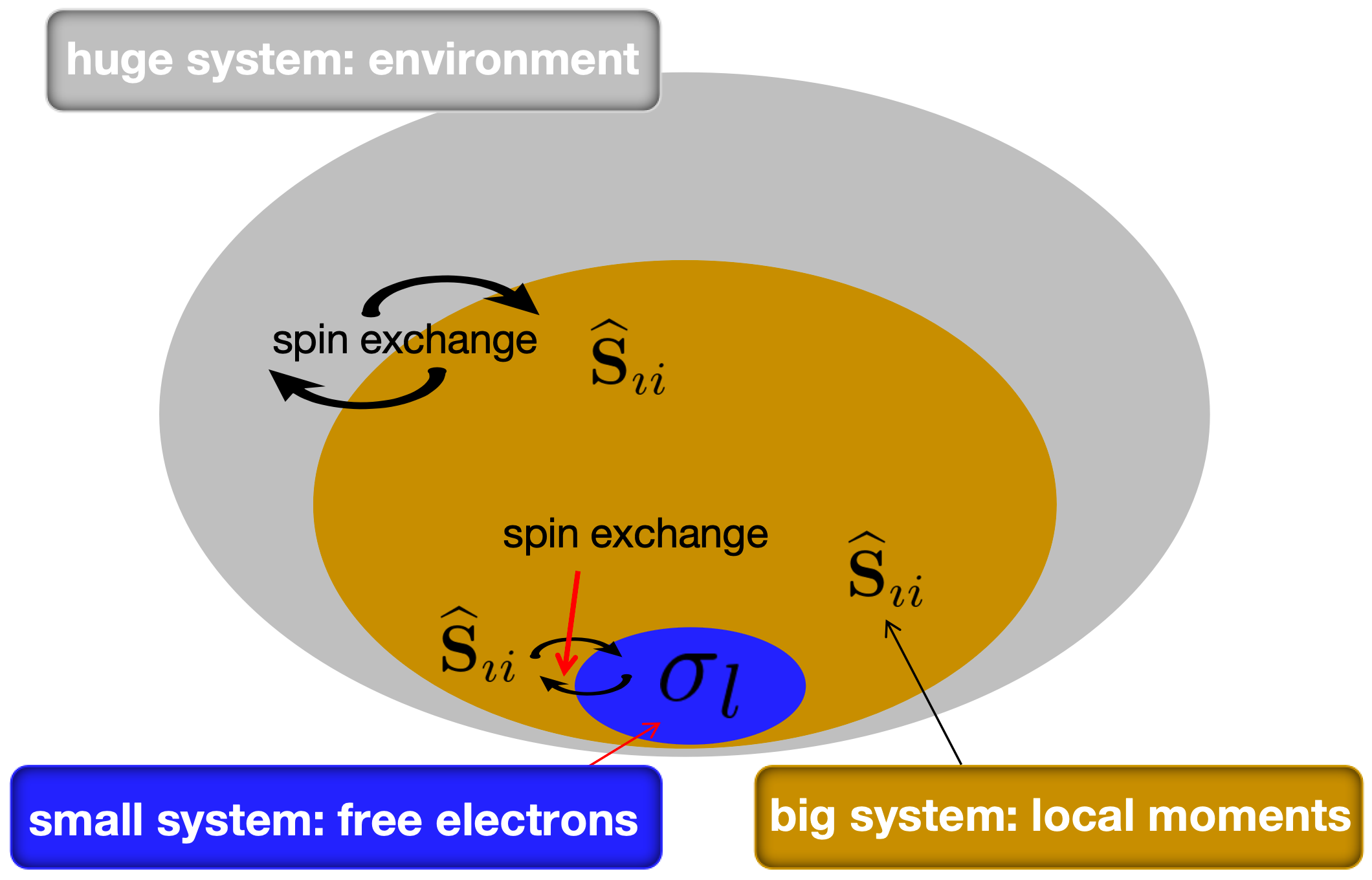} 
\end{center}
\caption{(Color online) The MR effect: An interacting and nonequilibrium problem in an open quantum system.}
\label{OPSaf}
\end{figure}

Before presenting our microscopic theory, we qualitatively describe the essence of the MR and the challenges it presents. Here, we investigate spin exchange in a composite system consisting of itinerant electrons and localized moments (see Fig.~\ref{OPSaf}). Thus, this is fundamentally a many-body problem. However, the composite system, in principle, will not reach a steady state on its own, as its dynamics are governed by a unitary Hamiltonian \eqref{dafvgsb}. During the spin exchange process, total spin is conserved; it merely transfers back and forth between the itinerant electrons and the localized moments, and thus no spin relaxation occurs. However, as is well known, such a composite system is not completely isolated - it is coupled to a huge environment and eventually reaches a steady state. Since our primary interest lies in the transport of the itinerant electrons, we assume that the local moments constitute a large system compared to the itinerant electrons, with their thermodynamic behavior determined by the external environment, such as the magnetic field and temperature. Therefore, the MR effect, by its nature, is an interacting and nonequilibrium problem in an open quantum system.

\section{Open quantum theory} \label{openquantumtheory}

In this section, we present the details of the open quantum theory of a hybrid system consisting of free electrons and local moments, with the latter being treated as a quantum bath (or reservoir). The full dynamics of the many-body density matrix $\check{\varrho}(t)$, governed by the spin-exchange coupling in Eq.~\eqref{spin-exchange}, are described by the Liouville-von Neumann equation in the interaction picture relative to the noninteracting Hamiltonian $\check{H}_0 = \hat{H}_e + \widehat{H}_m$:
\begin{equation}\label{favefvedlf}
\dfrac{\partial}{\partial t}\check{\varrho}(t)=\dfrac{i}{\hbar}[\check{\varrho}(t), \check{\mathcal{V}}_{\mathrm{em}}(t)],
\end{equation}
where the interaction-picture operator is defined as 
\begin{eqnarray}
 \check{\mathcal{O}}(t)=e^{+i\check{H}_0t/\hbar}O(t)e^{-i\check{H}_0t/\hbar}.
\end{eqnarray}
We solve Eq.~\eqref{favefvedlf} approximately using a time-convolutionless projection operator technique. Under the second-order Born-Markov approximation~\cite{breuer2002theory}, we obtain the master equation for the reduced electron density matrix $\hat{\varrho}_e(t) = \text{Tr}_m[\check{\varrho}(t)]$: 
\begin{eqnarray} \label{mNZE}
\dfrac{\partial}{\partial t} \hat{\varrho}_e(t)&=& \frac{i}{\hbar}\text{Tr}_m [\hat{\varrho}_e(t)\widehat{\varrho}_m(t), \check{\mathcal{V}}_{\mathrm{em}}(t)] \\
&-& \int_{t_0}^{t} \frac{d\tau}{\hbar^2} \text{Tr}_m [[ \hat{\varrho}_e(t) \widehat{\varrho}_m(t),\check{\mathcal{V}}_{\mathrm{em}}(\tau)],\check{\mathcal{V}}_{\mathrm{em}}(t)]. \notag
\end{eqnarray}
Here, $\widehat{\varrho}_m(t)=\text{Tr}_e[\check{\varrho}(t)]$ is the  reduced density matrix of the local moments, and $\text{Tr}_{e/m}$ traces out the free electron/local moment degree of freedom.
The first term on the right-hand side of Eq.~\eqref{mNZE} represents a correction to the effective Zeeman magnetic field experienced by the electrons due to the average spin expectation of the local moments. It can be compactly expressed as
\begin{eqnarray} 
  \hat{\mathcal{B}}_1(t)\hat{\varrho}_e(t) =-i \langle \widehat{L}_{n}(t)\rangle [ \hat{\varrho}_e(t) , \hat{\mathcal{V}}^{n}(t)],
\end{eqnarray}
where we have combined the indices into a single label $n=(\nu,\imath,i)$ for brevity. The second term in Eq.~\eqref{mNZE} acts as a collision integral, $\hat{\mathcal{J}}(\hat{\varrho}_e)$, which describes dissipation and decoherence. It is given by
\begin{eqnarray}
\hat{ \mathcal{J}}(\hat{\varrho}_e)&=&  \int_{t_0}^{t}  d\tau
 \left\lbrace  \mathcal{D}_{n_1n_2}(\tau-t) \left[\hat{\mathcal{V}}^{n_2}(t),\hat{\varrho}_e(t)\hat{\mathcal{V}}^{n_1}(\tau) \right]   \right.\notag\\
 &-& \left.  \mathcal{D}_{n_2n_1}(t-\tau)\left[ \hat{\mathcal{V}}^{n_2}(t),\hat{\mathcal{V}}^{n_1}(\tau) \hat{\varrho}_e(t) \right]  \right\rbrace.  
\end{eqnarray}
Here, $\mathcal{D}_{n_1n_2}(t_1-t_2)$ is the two-time correlation function of the quantum reservoir (i.e., local moments), defined as
\begin{align}
    \mathcal{D}_{n_1n_2}(t_1-t_2) = \langle \widehat{\mathcal{L}}_{n_1}(t_1)\widehat{\mathcal{L}}_{n_2}(t_2)\rangle.
\end{align}
For simplicity, we assume the magnetic order is determined solely by the Hamiltonian of the local moments, Eq. \eqref{bvabggb}. This treatment neglects the back-action of the itinerant electrons (the small subsystem) on the magnetic configuration of the local moments (the large reservoir), as illustrated in Fig.~\ref{OPSaf}.

Transforming into the Schrödinger picture, we obtain the master equation for the reduced electronic density matrix $\hat{\rho}(t)$ of free electrons 
\begin{equation}  \label{mSREPRE}
\dfrac{\partial}{\partial t} \hat{\rho}(t) - \frac{i}{\hbar} \left[\hat{\rho}(t), \hat{H}_e  \right]- \hat{B}_1(t)\hat{\rho}(t)=\hat{ J}(\hat{\rho}).
\end{equation} 
Now, we can clearly see that the term $\hat{B}_1(t)\hat{\rho}(t)$ is responsible for effective magnetic field experienced by the electrons due to the average spin expectation $\langle \widehat{L}_{n}(t)\rangle\propto \langle \widehat{S}_{n}(t)\rangle $ of the local moments 
\begin{eqnarray}  \label{mdjanv}
  \hat{B}_1(t)\hat{\rho}(t) =-i \langle \widehat{L}_{n}(t)\rangle [ \hat{\rho}(t) , \hat{V}^{n}],
\end{eqnarray}
The term $\hat{\mathcal{J}}(\hat{\rho})$ is a collision integral that describes spin-dependent electron scattering from the fluctuating local moments:
\begin{align} \label{mfsanfvfnv}
 \hat{ J}(\hat{\rho})&= \int_{0}^{t-t_0} d\tau  \left\lbrace  \mathcal{D}_{n_2n_1}(-\tau) \left[ \hat{V}^{n_1},\hat{\rho}(t)\hat{\mathcal{V}}^{n_2}(-\tau) \right]\right.\notag\\
 &-\left. \mathcal{D}_{n_1n_2}(+\tau)  \left[ \hat{V}^{n_1},\hat{\mathcal{V}}^{n_2}(-\tau) \hat{\rho}(t)\right]   \right\rbrace.  
\end{align}
with
\begin{align}
    \hat{\mathcal{V}}^{n_2}(-\tau) = e^{-i\hat{H}_e \tau/\hbar} \hat{V}^{n_2} e^{i\hat{H}_e \tau/\hbar}.
\end{align}
Here, the upper limit of the integral will be extended to infinity, consistent with the Markov approximation.
Hereafter, we focus on the quantum transport properties of the itinerant electrons. For the sake of brevity, we will omit the subscript $e$ on the electron density matrix $\hat{\rho}$.

\subsection{From many- to one-body master equation}

In this subsection, we derive a single-particle master equation from the many-body master equation for the electron density matrix $\hat{\rho}(t)$, given by Eq. \eqref{mSREPRE}. This reduction is essential as most quantum transport observables—such as charge and spin  densities and currents—are expressed in terms of single-particle operators. The central object is the single-particle density matrix, defined as
\begin{equation} \label{jgsufgnn}
    \varrho_{\alpha_2\alpha_1}(t)= \langle \alpha_2\vert \hat{\varrho}(t) \vert \alpha_1\rangle=\mathrm{Tr}_e\{\hat{f}^+_{\alpha_1}\hat{f}^{}_{\alpha_2}\hat{\rho}(t)\}.
\end{equation}
where $\alpha=(\vec{p},\sigma)$ labels states in an arbitrary orthonormal basis ${\vert\alpha\rangle}$ (e.g., spin-momentum states). Note that the index ordering in the single-particle density matrix is the conjugate of the ordering for the field operators inside the trace. The equation of motion for the single-particle density matrix is obtained by taking the trace of Eq. \eqref{mSREPRE}:
\begin{equation} \label{mRFE} 
\dfrac{\partial}{\partial t} \varrho_{\alpha_2\alpha_1}(t)-\frac{i}{\hbar}[\varrho(t), \hat{\mathcal{H}}^e_0]_{\alpha_2\alpha_1}-\frac{i}{\hbar}[\varrho(t), \hat{\mathcal{V}}_m]_{\alpha_2\alpha_1}=\mathcal{ J}_{\alpha_2\alpha_1}(\hat{\varrho}).
\end{equation} 
The effective mean-field potential arising from the local moments is given by
\begin{eqnarray} \label{bgabgfg}
  \hat{\mathcal{V}}_m =-\frac{1}{2}  N_{\imath}  \mathcal{J}^{\nu}_{\imath}   \langle \widehat{S}^{\nu}_{\imath}\rangle \hat{\sigma}^{\nu}  ,
\end{eqnarray}
which represents a renormalization of the electronic Zeeman energy.
For simplicity, we have worked in a plane-wave basis
\begin{align}
   \langle \vec{r} \vert\alpha\rangle=\frac{1}{\sqrt{V}} e^{\vec{k}\cdot\vec{r}} \vert\sigma\rangle.
\end{align}
Following the procedure detailed in Appendix~\ref{fvadkfvk}, the many-body collision integral (Eq. \ref{mfsanfvfnv}) reduces to a single-particle form as follows 
\begin{widetext}
\begin{align} \label{mfagagttgt1}
   \mathcal{J}^{\sigma_2\sigma_1}_{\vec{k}_2\vec{k}_1}(\check{\varrho})&=\frac{1}{V^2} \left\lbrace \mathcal{D}^-_{n_2n_1}(\omega^{\alpha_5}_{\alpha_6})\left[e^{-i(\vec{k}_2-\vec{k}_3)\cdot \vec{r}_{\vec{i}_1}-i(\vec{k}_5-\vec{k}_6)\cdot \vec{r}_{\vec{i}_2}} \sigma_{\sigma_2\sigma_3}^{\nu_1} \varrho^{\sigma_3\sigma_5}_{\vec{k}_3\vec{k}_5} \sigma_{\sigma_5\sigma_6}^{\nu_2}  \bar{\varrho}^{\sigma_6\sigma_1}_{\vec{k}_6\vec{k}_1}\right.\right.\\
    &-\left.e^{-i(\vec{k}_5-\vec{k}_6)\cdot \vec{r}_{\vec{i}_2}-i(\vec{k}_3-\vec{k}_1)\cdot \vec{r}_{\vec{i}_1}} \varrho^{\sigma_2\sigma_5}_{\vec{k}_2\vec{k}_5} \sigma_{\sigma_5\sigma_6}^{\nu_2}   \bar{\varrho}^{\sigma_6\sigma_3}_{\vec{k}_6\vec{k}_3} \sigma_{\sigma_3\sigma_1}^{\nu_1}\right] \notag \\
  &+\mathcal{D}^+_{n_1n_2}(\omega^{\alpha_6}_{\alpha_5})\left[e^{-i(\vec{k}_5-\vec{k}_6)\cdot \vec{r}_{\vec{i}_2}-i(\vec{k}_3-\vec{k}_1)\cdot \vec{r}_{\vec{i}_1}} \bar{\varrho}^{\sigma_2\sigma_5}_{\vec{k}_2\vec{k}_5}  \sigma_{\sigma_5\sigma_6}^{\nu_2}  \varrho^{\sigma_6\sigma_3}_{\vec{k}_6\vec{k}_3}   \sigma_{\sigma_3\sigma_1}^{\nu_1}\right.  \notag \\
  &-\left.\left. e^{-i(\vec{k}_2-\vec{k}_3)\cdot \vec{r}_{\vec{i}_1}-i(\vec{k}_5-\vec{k}_6)\cdot \vec{r}_{\vec{i}_2}}  \sigma_{\sigma_2\sigma_3}^{\nu_1}  \bar{\varrho}^{\sigma_3\sigma_5}_{\vec{k}_3\vec{k}_5} \sigma_{\sigma_5\sigma_6}^{\nu_2}    
   \varrho^{\sigma_6\sigma_1}_{\vec{k}_6\vec{k}_1}\right]\right\}, \notag 
\end{align}
\end{widetext}

with
\begin{align}
    \bar{\varrho}^{\sigma\sigma'}_{\vec{k}\vec{k}'}=\delta_{\vec{k}\vec{k}'}\delta_{\sigma\sigma'}-\varrho^{\sigma\sigma'}_{\vec{k}\vec{k}'}.
\end{align}
The appearance of the $\varrho\bar{\varrho}$ terms is a direct consequence of the fermionic statistics of the electrons, ensuring the Pauli exclusion principle is reflected in the scattering processes. Here, the off-diagonal components ($\sigma_2\neq\sigma_1$) of collision integral and density matrix are crucial for quantum decoherence theory, which contributes to a new extrinsic mechanism of anomalous Hall, spin Hall, and magnetoresistance effect~\cite{zhang2025anomalous,zhang2025theory}.

In this work, we focus on spin relaxation arising from the spin-exchange coupling. To this end, we restrict our analysis to the diagonal elements of the quantum kinetic equation (Eq. \ref{mRFE}), which govern the evolution of population numbers. Setting $\vec{k}_2=\vec{k}_1=\vec{k}$ and $\sigma_1=\sigma_2=\sigma$, the equation for the diagonal component of the density matrix, $\varrho^{\sigma}_{\vec{k}} \equiv \varrho^{\sigma\sigma}_{\vec{k}\vec{k}}$, simplifies to:
\begin{equation} 
\dfrac{\partial}{\partial t} \varrho^{\sigma}_{\vec{k}}(t)-\frac{i}{\hbar}[\varrho(t), \hat{\mathcal{H}}^e_0]^{\sigma\sigma}_{\vec{k}\vec{k}}-\frac{i}{\hbar}[\varrho(t), \hat{\mathcal{V}}_m]^{\sigma\sigma}_{\vec{k}\vec{k}}=\mathcal{ J}^{\sigma}_{\vec{k}}(\hat{\varrho}).
\end{equation} 
The corresponding collision integral for the diagonal elements is derived under the diagonal approximation for the density matrix 
\begin{align} \label{mfangunba}
    \varrho^{\sigma_2\sigma_1}_{\vec{k}_2\vec{k}_1}\simeq \delta_{\sigma_2\sigma_1}\delta_{\vec{k}_1\vec{k}_1}\varrho^{\sigma_1}_{\vec{k}_1},
\end{align}
and is given by:
\begin{widetext}
\begin{align} \label{rmfagagttgt1}
   \mathcal{J}^{\sigma}_{\vec{k}}(\check{\varrho})&=\frac{1}{V^2} \left\lbrace \left[\mathcal{D}^-_{n_2n_1}(\omega^{\vec{k}'\sigma'}_{\vec{k}\sigma})e^{-i(\vec{k}-\vec{k}')\cdot (\vec{r}_{\vec{i}_1}-\vec{r}_{\vec{i}_2})} \sigma_{\sigma\sigma'}^{\nu_1} \varrho^{\sigma'}_{\vec{k}'} \sigma_{\sigma'\sigma}^{\nu_2}  \bar{\varrho}^{\sigma}_{\vec{k}}\right.\right.\\
    &-\left.\mathcal{D}^-_{n_2n_1}(\omega^{\vec{k}\sigma}_{\vec{k}'\sigma'})e^{-i(\vec{k}-\vec{k}')\cdot (\vec{r}_{\vec{i}_2}-\vec{r}_{\vec{i}_1})} \varrho^{\sigma}_{\vec{k}} \sigma_{\sigma\sigma'}^{\nu_2}   \bar{\varrho}^{\sigma'}_{\vec{k}'} \sigma_{\sigma'\sigma}^{\nu_1}\right] \notag \\
  &+\left[\mathcal{D}^+_{n_1n_2}(\omega^{\vec{k}'\sigma'}_{\vec{k}\sigma})e^{-i(\vec{k}-\vec{k}')\cdot (\vec{r}_{\vec{i}_2}- \vec{r}_{\vec{i}_1})} \bar{\varrho}^{\sigma}_{\vec{k}}  \sigma_{\sigma\sigma'}^{\nu_2}  \varrho^{\sigma'}_{\vec{k}'}   \sigma_{\sigma'\sigma}^{\nu_1}\right.  \notag \\
  &-\left.\left. \mathcal{D}^+_{n_1n_2}(\omega^{\vec{k}\sigma}_{\vec{k}'\sigma'})e^{-i(\vec{k}-\vec{k}')\cdot (\vec{r}_{\vec{i}_1}-\vec{r}_{\vec{i}_2})}  \sigma_{\sigma\sigma'}^{\nu_1}  \bar{\varrho}^{\sigma'}_{\vec{k}'} \sigma_{\sigma'\sigma}^{\nu_2}    
   \varrho^{\sigma}_{\vec{k}}\right]\right\},\notag 
\end{align}
\end{widetext}
with
\begin{align} \label{fdkaknpm}
    \mathcal{D}^{+}_{n_1n_2}(\omega) =\int^{+\infty}_{0} d\tau e^{+i\omega\tau-\eta \tau}\mathcal{D}_{n_1n_2}(\tau),
\end{align}
\begin{align} \label{fdkaknnm}
    \mathcal{D}^{-}_{n_1n_2}(\omega) =\int^{0}_{-\infty} d\tau e^{+i\omega\tau+\eta \tau}\mathcal{D}_{n_1n_2}(\tau),
\end{align}
where $\sigma_{\sigma'\sigma}^{\nu}=\langle \sigma' \vert \sigma^{\nu}\vert \sigma'\rangle$ and $\bar{\varrho}^{\sigma}_{\vec{k}} = 1 - \varrho^{\sigma}_{\vec{k}}$ is the Pauli blocking factor. The collision integral $\mathcal{J}^{\sigma}_{\vec{k}}(\varrho)$ has a clear physical interpretation. The first and third terms (positive sign) describe scattering-in processes, where an electron transitions from a state $\vert \vec{k}'\sigma'\rangle$ to the state $\vert \vec{k}\sigma\rangle$, thereby increasing the population $\varrho^{\sigma}_{\vec{k}}$. Conversely, the second and fourth terms (negative sign) describe scattering-out processes, where an electron leaves the state $\vert \vec{k}\sigma\rangle$ for another state $\vert \vec{k}'\sigma'\rangle$, leading to a decrease in $\varrho^{\sigma}_{\vec{k}}$.

The primary challenge in evaluating the above collision integral lies in the summing over the positions of the local moments
\begin{align}
    \mathcal{D}^-_{n_2n_1}(\omega^{\vec{k}'\sigma'}_{\vec{k}\sigma})e^{-i(\vec{k}-\vec{k}')\cdot (\vec{r}_{\vec{i}_1}-\vec{r}_{\vec{i}_2})},
\end{align}
as the spin-spin correlation function $\mathcal{D}^{\pm}_{n_2n_1}(\omega^{\vec{k}'\sigma'}_{\vec{k}\sigma})$ depends on the relative locations $\vec{r}_{\vec{i}_1} - \vec{r}_{\vec{i}_2}$ of interacting moments [see Eq. \eqref{bvabggb}]. To simplify this interacting problem of local moments, we adopt the standard Weiss mean-field approximation.   That is, the interaction between local moments amounts only to a renormalization of the Zeeman term in the Hamiltonian~(\ref{bvabggb}),
\begin{eqnarray} \label{fvadfvdkfk}
g\mu _{B} \widehat{\vec{S}}_{\imath i}\cdot\vec{B}\to \textsl{g}\widehat{\vec{S}}_{\imath i}\cdot \vec{H}_{\imath i}.
\end{eqnarray}
The Weiss field $\vec{H}_{\imath i}$ acting on the  local moment $\widehat{\vec{S}}_{\imath i}$ becomes
\begin{equation} \label{fvdavfdf}
\vec{H}_{\imath i}=
\vec{B}
- \frac{1}{\textsl{g}\mu_B}J_{\imath i,\imath' i'}\langle \widehat{\vec{S}}_{\imath' i'}\rangle.
\end{equation}
This approximation dramatically simplifies the spin-spin correlations, reducing them to purely local terms: 
\begin{align} \label{gbagfbfg}
    \mathcal{D}^{\eta}_{n_2n_1}(\omega)\simeq \delta_{\vec{j}_2\vec{j}_1}\mathcal{D}^{\eta,\imath_1}_{\nu_2\nu_1}(\omega),
\end{align}
where $\mathcal{D}^{\eta,\imath}_{\nu_2\nu_1}(\omega)=\mathcal{D}^{\eta}_{\imath\nu_2,\imath\nu_1}(\omega)$. We now assume all moments within a given sublattice $\imath$ are equivalent, characterized by a uniform spin expectation value $\langle \text{S}_{\imath}^{\nu} \rangle$ and on-site correlation function $\langle  S_{\imath}^{\nu_2}(t) S_{\imath}^{\nu_1}(0) \rangle$. The Weiss field for a sublattice then becomes $\vec{H}_{\imath}=
\vec{B}
- \frac{1}{\textsl{g}\mu_B}J_{\imath i,\imath' i'}\langle \widehat{\vec{S}}_{\imath' i'}\rangle$. 
The spin expectation value and its autocorrelation are then determined self-consistently by this Weiss field~\cite{zhang2019theory}:
\begin{equation}\label{eqSavWeissexpl}
\langle S^{\imath}_{\parallel}\rangle = -S B_{S}\left(\beta S \epsilon^{\imath}_{L}\right),
\end{equation}
\begin{align} \label{spsp}
    \langle S^{\imath}_{\parallel}S^{\imath}_{\parallel} \rangle=S^{\imath}(S^{\imath}+1)+\coth(\beta \epsilon^{\imath}_{L}/2)\langle S_{\parallel} \rangle,
\end{align}
where $\beta=1/k_BT$, $B_{S}(x)$ is the Brillouin function, and $S^{\imath}_{\Vert }$ is the spin component in the direction of external the magnetic field $\vec{B}$. Crucially, the mean-field approximation renders the magnon dispersion flat, represented by a single effective Larmor energy $\epsilon^{\imath}_{L} = g\mu_B H_{\imath}$.  Applying the local correlation approximation \eqref{gbagfbfg}, the collision integral \eqref{rmfagagttgt1} simplifies significantly, factoring into a sum over independent sublattices:
\begin{align} \label{rmfagagfgabgttgt1}
   \mathcal{J}^{\sigma}_{\vec{k}}(\check{\varrho})&=  \frac{n_{\text{S}}^{\imath}}{V}\left\lbrace \left[\mathcal{D}^{-,\imath}_{ \nu_2\nu_1}(\omega^{\vec{k}'\sigma'}_{\vec{k}\sigma}) \sigma_{\sigma\sigma'}^{\nu_1} \varrho^{\sigma'}_{\vec{k}'} \sigma_{\sigma'\sigma}^{\nu_2}  \bar{\varrho}^{\sigma}_{\vec{k}}\right.\right.\\
    &-\left.\mathcal{D}^{-,\imath}_{\nu_2\nu_1}(\omega^{\vec{k}\sigma}_{\vec{k}'\sigma'}) \varrho^{\sigma}_{\vec{k}} \sigma_{\sigma\sigma'}^{\nu_2}   \bar{\varrho}^{\sigma'}_{\vec{k}'} \sigma_{\sigma'\sigma}^{\nu_1}\right] \notag \\
  &+\left[\mathcal{D}^{+,\imath}_{\nu_1\nu_2}(\omega^{\vec{k}'\sigma'}_{\vec{k}\sigma}) \bar{\varrho}^{\sigma}_{\vec{k}}  \sigma_{\sigma\sigma'}^{\nu_2}  \varrho^{\sigma'}_{\vec{k}'}   \sigma_{\sigma'\sigma}^{\nu_1}\right.  \notag \\
  &-\left.\left. \mathcal{D}^{+,\imath}_{\nu_1\nu_2}(\omega^{\vec{k}\sigma}_{\vec{k}'\sigma'})  \sigma_{\sigma\sigma'}^{\nu_1}  \bar{\varrho}^{\sigma'}_{\vec{k}'} \sigma_{\sigma'\sigma}^{\nu_2}    
   \varrho^{\sigma}_{\vec{k}}\right]\right\},\notag 
\end{align}
where $n_{\text{S}}^{\imath}=N_{\imath}/V$ is the density of local moments of $\imath$ sublattice.

In summary, the collision integral in Eq. \eqref{rmfagagfgabgttgt1} is derived based on three key approximations: (i) the second-order Born-Markov approximation applied to the many-body master equation [Eq. \eqref{mNZE}], (ii) the diagonal density matrix approximation for the electronic degrees of freedom [Eq. \eqref{mfangunba}], and (iii) the Weiss mean-field approximation for the local moments [Eq. \eqref{fvadfvdkfk}]. The mean-field treatment, in particular, greatly simplifies the many-body problem by reducing the spin-spin correlations to purely local functions [Eq. \eqref{gbagfbfg}]. It is important to note that the spin-spin correlation functions of the local moments, which enter the collision integral and are treated as a quantum bath, depend explicitly on temperature, external magnetic field strength, and the underlying magnetic configuration.

\subsection{Anisotropic spin relaxation}

In this subsection, we investigate the spin relaxation time for an arbitrary spin quantization axis, parameterized by the orientation angles $\theta \in [0, \pi]$ and $\phi \in [0, 2\pi)$. To facilitate this analysis, we introduce a rotated spin basis defined by the transformation:
\begin{align} 
\left|\Uparrow\right\rangle &= \cos \left(%
\frac{\theta}{2}\right)\left\vert \uparrow\right\rangle -e^{-i\phi }\sin \left(\frac{%
\theta }{2}\right)\left\vert \downarrow \right\rangle, \label{meqUpDnstsvphi1}\\
\left|\Downarrow\right\rangle &= \cos \left(%
\frac{\theta}{2}\right)\left\vert \downarrow \right\rangle+e^{+i\phi }\sin \left(\frac{%
\theta}{2}\right)\left\vert \uparrow\right\rangle, \label{meqUpDnstsvphi2}
\end{align}
where $\left\vert \uparrow\right\rangle$ and $\left\vert \downarrow\right\rangle$ are the eigenstates of the bare $\sigma_z$ operator (aligned parallel and antiparallel to the external magnetic field, respectively). The states $\left|\Uparrow\right\rangle$ and $\left|\Downarrow\right\rangle$ are the eigenstates of the rotated Pauli operator 
\begin{align}
    \sigma_{\mathbf{z}} = -\sin\theta\cos\phi\sigma_{x} + \sin\theta\sin\phi\sigma_{y} + \cos\theta\sigma_{z},
\end{align}
with eigenvalues $+1$ and $-1$, respectively. This transformation corresponds to a rotation of the spin quantization axis, implemented by the matrix $R$ which performs a rotation by angle $\phi$ about the $z$-axis followed by a rotation by angle $\theta$ about the $y$-axis. The spin operators for both the free electrons and the local moments transform under this rotation as:
\begin{align} \label{mRotationofspin}
\begin{bmatrix}
\hat{\sigma} _{\mathbf{x}} \\
\hat{\sigma} _{\mathbf{y}} \\
\hat{\sigma} _{\mathbf{z}}%
\end{bmatrix}%
 =
R
\begin{bmatrix}
\hat{\sigma} _{x} \\
\hat{\sigma} _{y} \\
\hat{\sigma} _{z}%
\end{bmatrix}%
 ,
 \begin{bmatrix}
 \widehat{L}^{i} _{\mathbf{x}} \\
 \widehat{L}^{i} _{\mathbf{y}} \\
 \widehat{L}^{i}_{\mathbf{z}}%
\end{bmatrix}%
 =
R
\begin{bmatrix}
 \widehat{L}^{i} _{x} \\
 \widehat{L}^{i} _{y} \\
 \widehat{L}^{i}_{z}%
\end{bmatrix}
,
\end{align}
where the rotation matrix is given by
\begin{align} \label{mRotation}
R =
\begin{bmatrix}
\cos \theta\cos \phi  & \mathtt{-}\cos \theta \sin \phi  &
\sin \theta  \\
\sin \phi  & \cos \phi  & 0 \\
\mathtt{-}\sin \theta\cos \phi & \sin \theta \sin \phi &
\cos \theta %
\end{bmatrix}
.
\end{align}%
Expressed in this rotated basis, the spin-exchange coupling term in Eq.~\eqref{spin-exchange} takes the form 
\begin{align} \label{meq:Vsdrs}
\check{V}_{\mathrm{em}}&=-\frac{1}{2}\sum_i\left\{\widehat{L}_{\mathbf{+}}^i\hat{\sigma}_{\mathbf{-}}(\vec{r}_i)+\widehat{L}_{\mathbf{-}}^i\hat{\sigma}_{\mathbf{+}}(\vec{r}_i)+\widehat{L}_{\mathbf{z}}^i\hat{\sigma}_{\mathbf{z}}(\vec{r}_i)\right\},
\end{align}
where the corresponding ladder operators are defined as 
\begin{align}
\widehat{L}^{i}_{\mathbf{+}}&=\widehat{L}^{i}_{\mathbf{x}}+i\widehat{L}^{i}_{\mathbf{y}},\hat{\sigma}_{\mathbf{+}}(\vec{r})=\frac{1}{2}[\hat{\sigma}_{\mathbf{x}}(\vec{r})+i\hat{\sigma}_{\mathbf{y}}(\vec{r})],\\
\widehat{L}^{i}_{\mathbf{-}}&=\widehat{L}^{i}_{\mathbf{x}}-i\widehat{L}^{i}_{\mathbf{y}},\hat{\sigma}_{\mathbf{-}}(\vec{r})=\frac{1}{2}[\hat{\sigma}_{\mathbf{x}}(\vec{r})-i\hat{\sigma}_{\mathbf{y}}(\vec{r})].
\end{align}

We now define the spin density along the arbitrary quantization axis defined by Eqs.~\eqref{meqUpDnstsvphi1} and \eqref{meqUpDnstsvphi2} as
\begin{align}
    N^{\sigma}=\frac{1}{V} \sum_{\vec{k}}\varrho^{\sigma}_{\vec{k}},
\end{align}
where $\sigma \in {\Uparrow, \Downarrow}$. The time evolution of this spin density is governed by  
\begin{align} \label{mfannunnrtt}
   \frac{\partial}{\partial t}  N^{\sigma} = \frac{1}{V} \sum_{\vec{k}}  \hat{\mathcal{J}}^{\sigma}_{\vec{k}}(\check{\varrho}),
\end{align}
where $\mathcal{J}^{\sigma}_{\vec{k}}(\varrho)$ is the collision integral given in Eq. \eqref{rmfagagfgabgttgt1}. We evaluate this expression in the ${\Uparrow, \Downarrow}$ basis for convenience. Substituting the explicit form of the collision integral, the rate equation for the spin density becomes 
\begin{align} \label{mdhfpdhaju}
  \frac{\partial}{\partial t}  N^{\sigma}=  W_{\sigma,\bar{\sigma}}(\vec{k},\vec{k}') \varrho^{\bar{\sigma}}_{\vec{k}'}  \bar{\varrho}^{\sigma}_{\vec{k}}-W_{\bar{\sigma},\sigma}(\vec{k}',\vec{k}) \varrho^{\sigma}_{\vec{k}}  \bar{\varrho}^{\bar{\sigma}}_{\vec{k}'} ,
\end{align}
where $\bar{\sigma}$ denotes the spin state opposite to $\sigma$. The transition rate $W_{\sigma\sigma'}(\vec{k}, \vec{k}')$ is given by
\begin{align}\label{mosdher}
    W_{\sigma,\sigma'}(\vec{k},\vec{k}')=\frac{1}{V^2} n_{\text{S}}^{\imath} \mathcal{D}^{\eta,\imath}_{\nu_2\nu_1}(\omega^{\sigma'}_{\vec{k}'}-\omega^{\sigma}_{\vec{k}}) \sigma_{\sigma\sigma'}^{\nu_1}  \sigma_{\sigma'\sigma}^{\nu_2}.
\end{align}
This form of the rate equation \eqref{mdhfpdhaju} has a clear physical interpretation: the first term represents the rate of scattering into state $|\vec{k} \sigma\rangle$ from all other states $|\vec{k}' \bar{\sigma}\rangle$ (gain term), while the second term represents scattering out of state $|\vec{k} \sigma\rangle$ (loss term).

In thermal equilibrium, we assume the electronic system is described by a Fermi-Dirac distribution, such that the density matrix is diagonal and given by
\begin{align} 
    \varrho^{\sigma}_{\vec{k}}=f(\epsilon^{\sigma}_{\vec{k}}),
\end{align}
where the spin-dependent energy is $\epsilon^{\sigma}_{\vec{k}}=\epsilon_{\vec{k}} + \epsilon^{\sigma}_B$. Here, $\epsilon^{\sigma}_B = \langle \sigma | \epsilon_B \hat{\sigma}_z | \sigma \rangle$ represents the Zeeman energy shift for the spin states $\sigma = {\Uparrow, \Downarrow}$ in the rotated basis. A crucial element of this derivation is the fluctuation-dissipation theorem, which relates the spin-spin correlation functions at positive and negative frequencies:
\begin{eqnarray} \label{mfluctuationdissipationtheorem}
  \mathcal{D}_{\nu_1\nu_2}(\epsilon)=e^{\beta \epsilon } \mathcal{D}_{\nu_2\nu_1}(-\epsilon).
  \end{eqnarray}
This relation directly implies that the scattering rates defined in Eq. \eqref{mosdher} satisfy the principle of detailed balance:
\begin{align} \label{farnur}
     W_{\sigma,\bar{\sigma}}(\vec{k},\vec{k}')=e^{\beta\left(\epsilon^{\bar{\sigma}}_{\vec{k}'}-\epsilon^{\sigma}_{\vec{k}}\right)}  W_{\bar{\sigma},\sigma}(\vec{k}',\vec{k}).
\end{align}
When the equilibrium form of the density matrix and the detailed balance condition \eqref{farnur} are inserted into the rate equation \eqref{mdhfpdhaju}, the gain and loss terms cancel exactly. Consequently, the time derivative of the spin density vanishes
\begin{align} \label{fannunnrtteq}
  \frac{\partial}{\partial t} N^{\sigma}=0,
\end{align}
as required for a system in thermal equilibrium. This result serves as an important consistency check for our formalism.

To describe out-of-equilibrium spin dynamics, we employ an ansatz that incorporates a spin-dependent chemical potential shift:
\begin{align}
    \varrho^{\sigma}_{\vec{k}}=f(\epsilon^{\sigma}_{\vec{k}}-\mu^{\sigma}).
\end{align}
The deviation of the distribution function from its equilibrium value is given by $ \delta  \varrho^{\sigma}_{\vec{k}} 
= \varrho^{\sigma}_{\vec{k}}-f(\epsilon^{\sigma}_{\vec{k}})$ reads:
\begin{equation} \label{Ans}
 \delta  \varrho^{\sigma}_{\vec{k}} \simeq -f'(\epsilon^{\sigma}_{\vec{k}}) \mu^{\sigma}=
\beta
f(\epsilon^{\sigma}_{\vec{k}})\bar{f}(\epsilon^{\sigma}_{\vec{k}})\mu^{\sigma}, 
\end{equation}
with $\bar{f}(\epsilon^{\sigma}_{\vec{k}})=1-f(\epsilon^{\sigma}_{\vec{k}})$. The resulting deviation in spin density is then 
\begin{align} \label{mfifagne}
    \delta N^{\sigma}&= \sum_{\vec{k}}\delta  \varrho^{\sigma}_{\vec{k}}= \beta
f(\epsilon^{\sigma}_{\vec{k}})\bar{f}(\epsilon^{\sigma}_{\vec{k}})\mu^{\sigma}=\nu^{\sigma}_F\mu^{\sigma},
\end{align}
which defines the effective density of states
\begin{align} \label{mahdadf}
    \nu^{\sigma}_F=\int d\epsilon
\nu(\epsilon)\beta
f(\epsilon^{\sigma}_{\vec{k}})\bar{f}(\epsilon^{\sigma}_{\vec{k}}).
\end{align}
At zero temperature, this simplifies to $\nu^{\sigma}_F = \nu(\epsilon_F - \epsilon^{\sigma}_B)$, the density of states at the spin-split Fermi energy. Using Eq.~\eqref{mfifagne}, we linearize the rate equation \eqref{mdhfpdhaju} to obtain the dynamics of the spin density deviation:
\begin{align} \label{dhdhaju}
  \frac{\partial}{\partial t} \delta N^{\sigma}&= \beta    W_{\sigma,\bar{\sigma}}(\vec{k},\vec{k}')  f(\omega^{\bar{\sigma}}_{\vec{k}'}) \bar{f}(\omega^{\sigma}_{\vec{k}})(\mu^{\bar{\sigma}}  -\mu^{\sigma}  ).
\end{align}

The out-of-equilibrium charge and spin densities along the quantization axis are defined, respectively, as 
\begin{align}
    \delta N^{o}&=\nu^{\Uparrow}_F\mu^{\Uparrow}+\nu^{\Downarrow}_F\mu^{\Downarrow},
\end{align}
\begin{align}
    \delta N^{\Vert}&=\nu^{\Uparrow}_F\mu^{\Uparrow}-\nu^{\Downarrow}_F\mu^{\Downarrow}.
\end{align}
The time evolution of the charge density is given by the sum of the spin-resolved rates  
\begin{align}
    \frac{\partial}{\partial t}\delta N^{o}=\frac{\partial}{\partial t} N^{\Uparrow}+\frac{\partial}{\partial t} N^{\Downarrow}=0,
\end{align}
which explicitly demonstrates charge conservation within our model. 
The evolution of the spin density is governed by 
\begin{align} 
   \frac{\partial}{\partial t}\delta N^{\Vert}&=-2\beta   W_{\Downarrow,\Uparrow}(\vec{k}',\vec{k})f(\omega^{\Uparrow}_{\vec{k}})\bar{f}(\omega^{\Downarrow}_{\vec{k}'})(\mu^{\Uparrow}  -\mu^{\Downarrow} ).
\end{align}
Expressing the chemical potential difference in terms of $\delta N^{c}$ and $\delta N^{s}$ using the inverted relations $\mu^{\Uparrow/\Downarrow} = (\delta N^{c} \pm \delta N^{s}) / (\nu^{\Uparrow}_F + \nu^{\Downarrow}_F)$ leads to a coupled equation of motion:
\begin{align} \label{faghtg}
   \frac{\partial}{\partial t}\delta N^{\Vert}&=-\frac{1}{T_{\Vert}}\delta N^{\Vert}-\frac{1}{T_{0}}\delta N^{0},
\end{align}
where the spin relaxation rate $(1/T_{\Vert})$ and the charge-spin conversion rate $(1/T_0)$ are given by
\begin{align} \label{mfajgatup}
    \frac{1}{T_{\Vert}}=\left(\frac{\beta}{\nu^{\Uparrow}_F}+\frac{\beta}{\nu^{\Downarrow}_F}\right)   W_{\Downarrow,\Uparrow}(\vec{k}_3,\vec{k})f(\epsilon^{\Uparrow}_{\vec{k}})\bar{f}(\epsilon^{\Downarrow}_{\vec{k}_3}),
\end{align}
\begin{align} \label{mfajgatup0}
    \frac{1}{T_{0}}=\left(\frac{\beta}{\nu^{\Uparrow}_F}-\frac{\beta}{\nu^{\Downarrow}_F}\right)   W_{\Downarrow,\Uparrow}(\vec{k}_3,\vec{k})f(\epsilon^{\Uparrow}_{\vec{k}})\bar{f}(\epsilon^{\Downarrow}_{\vec{k}_3}).
\end{align}
In the steady state ($\partial_t \delta N^s = 0$), Eq. \eqref{faghtg} yields a direct proportionality between the non-equilibrium spin density and charge density: 
\begin{align}
    \delta N^{\Vert}=-\frac{T_{\Vert}}{T_{0}}\delta N^{0}. 
\end{align}
This result signifies a mechanism for charge-to-spin conversion induced by the local moments, which is active whenever the spin-subband densities of states differ ($\nu^{\Uparrow}_F \neq \nu^{\Downarrow}_F$), such as in a finite external magnetic field. In the following section, we will derive explicit expressions for the longitudinal and transverse spin relaxation times from the general result in Eq. \eqref{mfajgatup}.

\subsubsection{Longitudinal spin relaxation time}

We now evaluate the longitudinal spin relaxation time, which corresponds to the case where the quantization axis is parallel to the external magnetic field [$\theta = 0$ in Eq. \eqref{mRotation}]. In this configuration, the rotated ladder operators for the local moments simplify to: 
\begin{align} \label{msjaglkk1}
L^{\imath}_{\mathbf{+}}&=e^{+i\phi}\left[\frac{\mathcal{J}_{\imath}^x+\mathcal{J}_{\imath}^y}{2}S^{\imath}_++\frac{\mathcal{J}_{\imath}^x-\mathcal{J}_{\imath}^y}{2}S^{\imath}_-\right],
\end{align}
\begin{align}\label{msjaglkk2}
L^{\imath}_{\mathbf{-}}&=e^{-i\phi}\left[\frac{\mathcal{J}_{\imath}^x-\mathcal{J}_{\imath}^y}{2}S^{\imath}_++\frac{\mathcal{J}_{\imath}^x+\mathcal{J}_{\imath}^y}{2}S^{\imath}_-\right].
\end{align}
A crucial consequence of anisotropic spin-exchange coupling ($\mathcal{J}_{\imath}^x \neq \mathcal{J}_{\imath}^y$) is the breaking of spin conservation. The $\sigma_{+}L^{\imath}_{\mathbf{-}}$ term, for instance, demonstrates that a spin-flip of the free electron ($\sigma_+$) can now be accompanied by a spin-flip of the local moment in the same direction ($S^{\imath}_+$), a process forbidden in the isotropic case. Substituting Eqs. \eqref{msjaglkk1} and \eqref{msjaglkk2} into the scattering rate \eqref{mosdher} yields: 
\begin{widetext}
\begin{align} \label{mfpdnutgt}
    W_{\Downarrow,\Uparrow}(\vec{k}',\vec{k})&=\frac{\pi}{8\hbar^2V^2} n_{\text{S}}^{\imath} (\mathcal{J}_{\imath}^x+\mathcal{J}_{\imath}^y)^2  \left\langle S^{\imath}_{-}S^{\imath}_{+} \right\rangle (\omega^{\Uparrow}_{\vec{k}}-\omega^{\Downarrow}_{\vec{k}'}-\omega_L^{\imath})\\
    &+\frac{\pi}{8\hbar^2V^2} n_{\text{S}}^{\imath} (\mathcal{J}_{\imath}^x-\mathcal{J}_{\imath}^y)^2  \left\langle S^{\imath}_{+}S^{\imath}_{-} \right\rangle (\omega^{\Uparrow}_{\vec{k}}-\omega^{\Downarrow}_{\vec{k}'}+\omega_L^{\imath}).\notag 
\end{align}    
\end{widetext}
The first and second terms correspond to spin-conserving and non-conserving scattering processes, respectively. The spin-spin correlation functions governing these flip processes are determined by the parallel spin components:
\begin{align} \label{meqDDdeft11} 
    \left\langle S^{\imath}_{-}S^{\imath}_{+} \right\rangle(\omega)
&=\left[S^{\imath}(S^{\imath}+1)-\langle S^{\imath}_{\Vert
}S^{\imath}_{\Vert
}\rangle-\langle S^{\imath}_{\Vert
}\rangle\right]\delta(\omega- \omega^{\imath}_L),   
\end{align}
\begin{align} \label{meqDDdeft12}
\langle S^{\imath}_{+}S^{\imath}_{-} \rangle(\omega)
&=\left[S^{\imath}(S^{\imath}+1)-\langle S^{\imath}_{\Vert
}S^{\imath}_{\Vert
}\rangle+\langle S^{\imath}_{\Vert
}\rangle\right]\delta(\omega+ \omega^{\imath}_L),
\end{align}
where we have used the identity
\begin{align}
    S^{\imath}_{x}S^{\imath}_{x
}+S^{\imath}_{y}S^{\imath}_{y
}+S^{\imath}_{z}S^{\imath}_{z
}=S^{\imath}(S^{\imath}+1).
\end{align}
The delta functions enforce energy conservation during the scattering. Within the Weiss mean-field framework, the final expression for the longitudinal spin relaxation rate is 
\begin{widetext}
\begin{align} \label{mLTdppaAdU}
\frac{1}{\tau_\parallel}&=
\frac{\pi}{4\hbar}   n_{\text{S}}^{\imath}  \nu_F (\mathcal{J}_{\imath}^x\mathcal{J}_{\imath}^x+\mathcal{J}_{\imath}^y\mathcal{J}_{\imath}^y) \beta\epsilon_L^{\imath}
 n_{B}\left(\epsilon^{\imath}_L\right)[S^{\imath}(S^{\imath}+1)-\langle S^{\imath}_{\Vert
}S^{\imath}_{\Vert
}\rangle-\langle S^{\imath}_{\Vert
}\rangle]\\
&+\frac{\pi}{4\hbar}    n_{\text{S}}^{\imath} \nu_F (\mathcal{J}_{\imath}^x-\mathcal{J}_{\imath}^y)^2 \beta\epsilon_L^{\imath} [1+n_B(\epsilon_L^{\imath})] [S^{\imath}(S^{\imath}+1)-\langle S^{\imath}_{\Vert
}S^{\imath}_{\Vert
}\rangle+\langle S^{\imath}_{\Vert
}\rangle],  \notag 
\end{align}    
\end{widetext}
where the effective density of states is
\begin{align} \label{fdvafvdf}
    \nu_F=\left(\frac{1}{2\nu^{\Uparrow}_F}+\frac{1}{2\nu^{\Downarrow}_F}\right)\nu^{\Uparrow}_F\nu^{\Downarrow}_F,
\end{align}
Here, $n_B(\epsilon^{\imath}_L) = 1/(e^{\beta\epsilon^{\imath}_L} - 1)$ is the Bose-Einstein distribution function. The appearance of $n_B$ and $1+n_B$, derived microscopically from the many-body density matrix, reflects the bosonic nature of the local moments treated as a quantum bath. It is noteworthy that only spin-flip processes contribute to the longitudinal spin relaxation.

\subsubsection{Transverse spin relaxation time}

We now evaluate the transverse spin relaxation time, which corresponds to the quantization axis lying perpendicular to the external magnetic field [$\theta = \pi/2$ in Eq. \eqref{mRotation}]. In this configuration, the rotated ladder operators for the local moments are given by  
\begin{align} \label{msjadglkk1}
L^{\imath}_{\mathbf{+}}&=\mathcal{J}_{\imath}^zS^{\imath}_z+i\left[\frac{\sin\phi \mathcal{J}_{\imath}^x-i\cos\phi \mathcal{J}_{\imath}^y}{2}S^{\imath}_+\right.\\
&+\left.\frac{\sin\phi \mathcal{J}_{\imath}^x+i\cos\phi \mathcal{J}_{\imath}^y}{2}S^{\imath}_-\right],\notag 
\end{align}
\begin{align}\label{msjagdlkk2}
L^{\imath}_{\mathbf{-}}&= \mathcal{J}_{\imath}^zS^{\imath}_z-i\left[\frac{\sin\phi \mathcal{J}_{\imath}^x-i\cos\phi \mathcal{J}_{\imath}^y}{2}S^{\imath}_+\right.\\
&+\left.\frac{\sin\phi \mathcal{J}_{\imath}^x+i\cos\phi \mathcal{J}_{\imath}^y}{2}S^{\imath}_-\right].\notag 
\end{align}
A key distinction from the longitudinal case (Eqs. \eqref{msjaglkk1} and \eqref{msjaglkk2}) is the presence of a spin-conserving term proportional to $S^{\imath}_z$. This term will contribute to spin dephasing. Substituting Eqs. \eqref{msjadglkk1} and \eqref{msjagdlkk2} into the scattering rate \eqref{mosdher} yields three distinct physical contributions: 
\begin{widetext}
\begin{align} \label{mfpdnudaptgt}
    W_{\Downarrow,\Uparrow}(\vec{k}',\vec{k})&=\frac{\pi}{8\hbar^2V^2} n_{\text{S}}^{\imath} (\sin^2\phi \mathcal{J}_{\imath}^x\mathcal{J}_{\imath}^x+\cos^2\phi\mathcal{J}_{\imath}^y\mathcal{J}_{\imath}^y )  \left\langle S^{\imath}_{-}S^{\imath}_{+} \right\rangle (\omega^{\Uparrow}_{\vec{k}}-\omega^{\Downarrow}_{\vec{k}'}-\omega_L^{\imath})\\
    &+\frac{\pi}{8\hbar^2V^2} n_{\text{S}}^{\imath} (\sin^2\phi \mathcal{J}_{\imath}^x\mathcal{J}_{\imath}^x+\cos^2\phi \mathcal{J}_{\imath}^y\mathcal{J}_{\imath}^y ) \left\langle S^{\imath}_{+}S^{\imath}_{-} \right\rangle (\omega^{\Uparrow}_{\vec{k}}-\omega^{\Downarrow}_{\vec{k}'}+\omega_L^{\imath})\notag \\
    &+\frac{\pi}{2\hbar^2V^2} n_{\text{S}}^{\imath} \mathcal{J}_{\imath}^z\mathcal{J}_{\imath}^z  \langle S^{\imath}_{\Vert}S^{\imath}_{\Vert} \rangle (\omega^{\Uparrow}_{\vec{k}}-\omega^{\Downarrow}_{\vec{k}'}).\notag
\end{align}    
\end{widetext}
The first two terms represent spin-flip processes, while the third term represents a spin-conserving, dephasing process. The final expression for the transverse spin relaxation rate within the Weiss mean-field framework is 
\begin{widetext}
\begin{align} \label{mfjanwegn}
    \frac{1}{\tau_{\perp}}&=\frac{\pi}{4\hbar}    n_{\text{S}}^{\imath} \nu_F (\sin^2\phi \mathcal{J}_{\imath}^x\mathcal{J}_{\imath}^x+\cos^2\phi \mathcal{J}_{\imath}^y\mathcal{J}_{\imath}^y )  \beta\epsilon_L^{\imath} n_B(\epsilon_L^{\imath}) \left\langle S^{\imath}_{-}S^{\imath}_{+} \right\rangle\\
    &+\frac{\pi}{4\hbar}    n_{\text{S}}^{\imath}\nu_F (\sin^2\phi \mathcal{J}_{\imath}^x\mathcal{J}_{\imath}^x+\cos^2\phi \mathcal{J}_{\imath}^y\mathcal{J}_{\imath}^y )  \beta\epsilon_L^{\imath} [1+n_B(\epsilon_L^{\imath})] \left\langle S^{\imath}_{+}S^{\imath}_{-} \right\rangle\notag \\
    &+\frac{\pi}{\hbar} n_{\text{S}}^{\imath} \mathcal{J}_{\imath}^z\mathcal{J}_{\imath}^z  \langle S^{\imath}_{\Vert}S^{\imath}_{\Vert} \rangle .\notag
\end{align}
\end{widetext}
In contrast to the longitudinal case, the transverse relaxation is governed by both spin-flip processes (the first two terms) and spin-conserving dephasing processes (the third term). The latter, which persists even in the absence of spin-flip scattering, is responsible for the pure dephasing of transverse spin components.

In summary, the analytical results presented in this work rely on several key approximations. First, to isolate the relaxation mechanisms stemming purely from the spin-exchange coupling, we have assumed a spin-independent density of states at the Fermi energy. This excludes contributions from spin-resolved band structure effects. Second, we have decoupled the spin-exchange coupling from spin-orbit coupling to specifically calculate the spin relaxation time arising from the local moments. Third, we disregard the feedback effect of the itinerant electrons on the thermodynamic equilibrium of the local moment system. The expressions for the spin relaxation times in Eqs. \eqref{mLTdppaAdU} and \eqref{mfjanwegn} are derived within the combined framework of the Born-Markov and Weiss mean-field approximations, yielding a compact and analytical form. Finally, we note that the spin expectation values and spin-spin correlation functions that enter these expressions are intrinsically dependent on the magnetic configuration of the local moments.

\section{Anisotropic spin diffusion equation} \label{diffusionequation}

With the longitudinal and transverse spin relaxation times established in Eqs. \eqref{mLTdppaAdU} and \eqref{mfjanwegn}, we can now construct the spin density continuity equation in the rotated basis. This equation describes the evolution of the spin chemical potential vector $\boldsymbol{\mu}_s = (\mu^{\textbf{x}}_s, \mu^{\textbf{y}}_s, \mu^{\textbf{z}}_s)$ 
\begin{align}
    \frac{d}{dt}
    \begin{bmatrix}
        \mu^{\textbf{x}}_s\\
        \mu^{\textbf{y}}_s\\
        \mu^{\textbf{z}}_s
    \end{bmatrix}+\nabla\cdot \begin{bmatrix}
        j^{\textbf{x}}_s\\
        j^{\textbf{y}}_s\\
        j^{\textbf{z}}_s
    \end{bmatrix}= -
\begin{bmatrix}
\tau^{-1}_{\perp}  & 0 &
-\omega_L  \\
0  & \tau^{-1}_{\perp}  & +\omega_L \\
+\omega_L & -\omega_L &
\tau^{-1}_{\Vert}
\end{bmatrix}
\begin{bmatrix}
        \mu^{\textbf{x}}_s\\
        \mu^{\textbf{y}}_s\\
        \mu^{\textbf{z}}_s
\end{bmatrix}.
\end{align}
The spin current is governed by Fick's law as follows 
\begin{align} \label{davfk}
    j^{\nu}_s=-\mathcal{D}\nabla \mu^{\nu}_s,
\end{align}
where the diffusion coefficient $\mathcal{D}$ is assumed to be isotropic for simplicity. As indicated by Eq. \eqref{bgabgfg}, the first-order correction from the spin-exchange coupling renormalizes the Hanle spin precession frequency. This introduces an effective spin-exchange field, shifting the frequency to  
\begin{align} \label{fvdkvkakf}
   \omega
_{L}=\omega_{B}-\frac{1}{\hbar}\sum_{\imath}n^{\imath}_{
\mathrm{S}}\mathcal{J}^{\imath}_{sd}\langle S^{\imath}_{\Vert }\rangle.
\end{align}
To express the equations in the original laboratory frame, we apply the inverse of the rotation matrix $R$ defined in Eq. \eqref{mRotation}. The spin chemical potential in the rotated basis relates to the original components via
\begin{align}
    \begin{bmatrix}
        \mu^{\textbf{x}}_s\\
        \mu^{\textbf{y}}_s\\
        \mu^{\textbf{z}}_s
\end{bmatrix}=R\begin{bmatrix}
        \mu^{x}_s\\
        \mu^{y}_s\\
        \mu^{z}_s
\end{bmatrix}=\left[\begin{array}{ccc}
* & * & * \\
* & * & * \\
\hat{b}_x & \hat{b}_y & \hat{b}_z
\end{array}\right]
\begin{bmatrix}
        \mu^{x}_s\\
        \mu^{y}_s\\
        \mu^{z}_s
\end{bmatrix},
\end{align}
where the third row of $R$ is the unit vector $\hat{\vec{b}}$, referring to the quantization axis, i.e., $\hat{\vec{b}}=[\mathtt{-}\sin \theta\cos \phi, \sin \theta \sin \phi,
\cos \theta]$.  Transforming the continuity equation back to the lab frame and considering the steady state ($d/dt = 0$), we obtain 
\begin{align} \label{mfvdvfvdf}
   \nabla\cdot j^{\nu}_s=-\tau^{-1}_{\perp}\mu^{\nu}_s-(\tau^{-1}_{\Vert}-\tau^{-1}_{\perp})\hat{b}_{\nu}\hat{b}_{\kappa}\mu^{\delta}_s-\omega_L\epsilon_{\nu\kappa\delta}\mu^{\delta}_s,
\end{align}
where $\nu, \kappa, \delta = {x, y, z}$ and $\epsilon_{\nu\kappa\delta}$ is the Levi-Civita symbol. We now consider a mesoscopic magnetic material that is translationally invariant in the $\hat{x}$-$\hat{y}$ plane, such that the spin accumulation $\mu_{s}^{\nu}$ depends only on the coordinate $z$. Substituting the spin current expression \eqref{davfk} into Eq. \eqref{mfvdvfvdf} yields a diffusion equation with anisotropic spin relaxation  
\begin{equation} \label{3.3}
\partial _{z}^{2}\mu _{\mathrm{s}}^{\nu}=\ell _{\mathrm{\bot }}^{-2}\delta
_{\nu \kappa }\mu _{\mathrm{s}}^{\kappa }+\left( \ell _{\mathrm{\parallel }%
}^{-2}-\ell _{\mathrm{\bot }}^{-2}\right) \hat{b}_{\nu}\hat{b}_{\kappa}\mu _{\mathrm{s}%
}^{\kappa} 
-\ell _{\mathrm{L}}^{-2}\epsilon _{\nu \kappa \delta }\hat{b}_{\delta }\mu _{%
\mathrm{s}}^{\delta},  
\end{equation}%
The  Hanle spin precession length is given by $\ell _{\mathrm{L}}=\sqrt{\mathcal{D}/\omega _{L}}$, while the longitudinal and transverse spin diffusion lengths are expressed as $\ell _{\mathrm{\bot }}=\sqrt{\mathcal{D}\tau _{\mathrm{\bot }}}$ and $%
\ell _{\mathrm{\parallel }}=\sqrt{\mathcal{D}\tau _{\mathrm{\parallel }}}$, respectively.

\begin{figure}[t!]
\begin{center}
\includegraphics[width=0.47\textwidth]{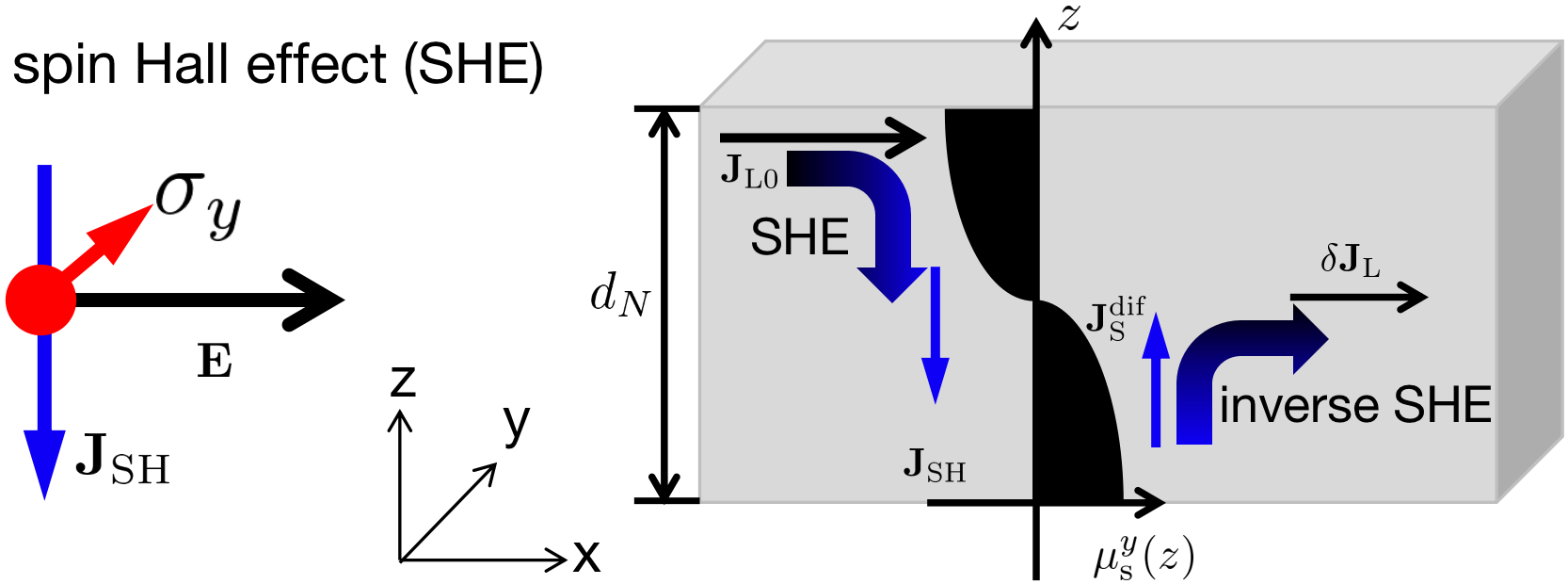} 
\end{center}
\caption{(Color online) The MR effect  arises from a two-step charge–spin conversion process in a monolayer ferromagnetic material. The left panel plots the spin Hall effect, where electric field ($\vec{E}$), the spin polarization ($\sigma_y$) and flowing direction ($\hat{z}$) of spin Hall current ($\vec{J}_{\text{SH}}$) are perpendicular to each other.}
\label{STORY}
\end{figure}

Besides the spin-exchange coupling~\eqref{spin-exchange}, another dominant ingredient of MR is the spin-orbit coupling~\cite{mcguire1975anisotropic}.  The MR stems from the combination of the spin Hall effect and its inverse effect that arise from the spin-orbit coupling~\cite{maekawa2017spin,sinova2015spin}, as depicted in Fig.~\ref{STORY}. In the first step charge current $\vec{J}_{\text{L0}}=\sigma_{\text{D}}\vec{E}$ is converted to a drift spin current $\vec{J}_{\text{SH}}=\theta_{\text{SH}}(\hat{y}\times \vec{J}_{\text{L0}})$ via the spin Hall effect, as shown by the left-curved arrow, Here, $\theta_{\text{SH}}\propto V_{\text{so}}$ represents the charge-spin conversion ratio of the spin Hall effect~\cite{sinova2015spin,zhang2024altermagnet}, $\sigma_{\text{D}}$ is the Drude conductivity, and $\vec{E}$ is an electric field in $x$-axis direction. The drift spin current polarized in $y$-axis direction and flowing in $z$-axis direction leads to considerable spin accumulation $\mu
_{\mathrm{s}}^{y}(z)$ that accounts for the diffusive spin current, $J^{\text{dif}}_{\text{S}}=-\frac{\sigma_{\text{D}} }{2e}\partial _{z}\mu
_{\mathrm{s}}^{y}(z)$. The boundary conditions require that the total spin
current vanishes at the top ($z=d_N$) and bottom ($z=0$) interfaces~\cite{chen2013theory,velez2016hanle}
\begin{align} \label{mfvavkdfmv}
    \left. J_{\mathrm{s}}^{\nu }\right\vert_{z=0,d_{N}}=-\frac{\sigma_{\text{D}} }{2e}\left. \partial _{z}\mu
_{\mathrm{s}}^{\nu }\right\vert _{z=0,d_{N}}+\delta _{\nu y}\theta _{\mathrm{SH}}\sigma_{\text{D}}E=0,
\end{align}
with $\nu =\{x,y,z\}$ where $d_N$ is the thickness of the ferromagnetic monolayer. In the second step, the diffusive spin current is converted back to charge current by the inverse spin Hall effect (right-curved arrow in Fig.~\ref{STORY}). Overall, both spin-exchange coupling and spin-orbit coupling are crucial for the MR and participate in diffusion equation  (\ref{3.3}) and boundary condition~\eqref{mfvavkdfmv}, respectively. Next, we quantitatively investigate the MR in ferromagnet (Sec.~\ref{ferromagneticMR}) and antiferromagnet (Sec.~\ref{antiferromagneticMR}).

\section{Results and Discussions} \label{resultsanddiscussions}

\subsection{Ferromagnetic MR} \label{ferromagneticMR}

Let us begin with ferromagnetic MR with one sublattice with $n_{
\mathrm{S}}=n^{\imath}_{
\mathrm{S}}$, $\langle S^{\nu}_{}\rangle=\langle S^{\nu}_{\imath}\rangle$, and $\langle \text{S}_{}^{\nu_2} \text{S}_{}^{\nu_1} \rangle = \langle \text{S}_{\imath i}^{\nu_2} \text{S}_{\imath i}^{\nu_1} \rangle$. There exist two crucial effects caused by the spin-exchange coupling \eqref{spin-exchange}, which have been overlooked by the previous theories of MR in a monolayer ferromagnetic material. First, the spin-exchange coupling shifts the Hanle spin precession frequency $\omega
_{L}=\omega_{B}-\textsl{g}\mu_B\mathcal{B}_{sd}$ [i.e. Eq.~\eqref{fvdkvkakf}] by spin-exchange field as follows 
\begin{equation} \label{elf}
 \mathcal{B}_{sd}=n_{
\mathrm{S}}\mathcal{J}\langle S_{\Vert }\rangle/(\hbar \textsl{g}\mu_B),
\end{equation}
with $\omega_B=\textsl{g}\mu _{B}B$, where we have assumed isotropic spin-exchange coupling ($\mathcal{J}^{\nu}_{\imath}=\mathcal{J}$) for simplicity, $n_{\mathrm{S}}$ is the density of local moments, and $S_{\Vert }$ is the spin component in the direction of the magnetization, which is assumed to be collinear to $\vec{B}$~\footnote{This is a widely used approximation for simplicity when we deal with the MR irrelevant to hysteresis and magnetostatic and magneto-crystalline anisotropy. Of course, following our method, we can directly include these effects by considering non-collinear case.}. Second, spin-exchange coupling causes anisotropic spin relaxation~\cite{zhang2019theory}. The longitudinal (transverse) spin
relaxation times $\tau _{\Vert }$ ($\tau _{\bot }$) can be expressed by the spin expectation value $\langle S_{\parallel}\rangle$ and spin-spin correlation function $\langle S^2_{\Vert
}\rangle$ of local moments as follows
\begin{align} \label{LongiSRT}
\frac{1}{\tau_{\Vert}}&= \frac{1}{\tau^{}_0}
+\frac{\pi}{\hbar}   n_{\text{S}}  \nu_F \mathcal{J}^2 \beta\epsilon_L
 n_{B}(\epsilon_L)[S(S+1)-\langle S^2_{\Vert
}\rangle-\langle S_{\Vert
}\rangle],
\end{align}
\begin{align} \label{TransSRT}
\frac{1}{\tau_{\perp}}&= \frac{1}{2\tau^{}_0}+\frac{1}{2\tau_{\Vert}}
+\frac{\pi}{\hbar}   n_{\text{S}}  \nu_F \mathcal{J}^2\langle S^2_{\Vert
}\rangle.
\end{align}
Here, $\nu_F$ is the density of state for each spin at Fermi energy  [see the expression of $\nu_F$ in Eq.~\eqref{fdvafvdf}],  that captures energy-resolved contributions to spin decoherence (i.e., spin relaxation and dephasing) and is assumed to be the same for spin-up and spin-down species. \textit{\textquotedblleft The density of states is different for majority and minority
spins"} is the necessary condition for the previous MR theory based on momentum relaxation time that strongly relies on the density of state. In our theory, focusing on the anisotropic spin relaxation time, the density of states for spin-up and spin-down species can be the same. The second term of Eq.~\eqref{LongiSRT} describes the spin-flip rate associated with magnon emission and absorption, while the third term of Eq.~\eqref{TransSRT} represents the spin dephasing rate arising from scattering processes during which the electron spin, being in a superposition of
spin-up and spin-down, acquires a precession phase about the $\vec{m}$-direction. $n_B(\epsilon_L)=1/(e^{\beta\epsilon_L}-1)$ is the Bose-Einstein distribution function at temperature $\beta=1/(k_BT)$. The Weiss mean field approximation simplifies the magnon dispersion as a constant  effective Larmor frequency 
$\epsilon_{L}=\hbar\omega_B-\langle S_{\Vert}\rangle\sum_{i} J_{ij} $, where $k_B$ is the Boltzmann constant.  The spin-spin correlation function can be expressed by   $\langle S^2_{\parallel} \rangle=S(S+1)+\coth(\beta\epsilon_{L}/2)\left\langle S_{\parallel} \right\rangle$~\cite{zhang2019theory}.

The first term, $\tau_0^{-1}$, denotes the isotropic spin relaxation rate that is unrelated to local moments and is assumed to be independent of both $B$ and $T$. The spin relaxation originates from spin-flip scattering mediated by a spin-flip potential ($V_{\text{sf}}$), whose spin-flip matrix element between momentum states $\mathbf{k}$ and $\mathbf{k}'$ is given by $\langle \mathbf{k}\uparrow \vert V_{\text{sf}} \vert \mathbf{k}'\downarrow \rangle$. The corresponding spin-flip scattering rate is proportional to the squared modulus of this matrix element, $\tau_0^{-1} \propto \vert\langle \mathbf{k}\uparrow \vert V_{\text{sf}} \vert \mathbf{k}'\downarrow \rangle\vert^2$. 
Such spin-flip processes may arise from extrinsic spin–orbit coupling or from spin-dependent static disorder. As a concrete example, we consider extrinsic spin–orbit coupling ($V_{\text{sf}} = V_{\text{so}}$), introduced by randomly distributed impurities located at positions $\mathbf{R}_j$, each described by a short-range potential $V_{\text{so}}(\mathbf{r}) = U \sum_j \delta(\mathbf{r} - \mathbf{R}_j)$, where $U$ characterizes the scattering strength. In momentum space, the corresponding spin–orbit-modified scattering potential takes the form $\sum_{\vec{k}\vec{k}'}[U^{o}_{\vec{k}\vec{k}'}-iU^{s}_{\vec{k}\vec{k}'}\vec{\sigma} \cdot (\vec{k} \times \vec{k}' )]$, 
where $U^{s}_{\mathbf{k}\mathbf{k}'} = l_{\text{so}}^{2} U \sum_j e^{-i(\mathbf{k} - \mathbf{k}')\cdot \mathbf{R}_j}$ represents the spin–orbit-dependent part, and $U^{o}_{\mathbf{k}\mathbf{k}'} = U \sum_j e^{-i(\mathbf{k} - \mathbf{k}')\cdot \mathbf{R}_j}$ is an artificially introduced spin-independent term~\cite{zhang2024altermagnet}. The resulting isotropic spin relaxation rate is then obtained as $\tau_0^{-1} \simeq \frac{8\pi}{3\hbar} n_i \nu_F k_F^4 l_{\text{so}}^4 U^2$
where $l_{\text{so}} = \hbar / (2mc)$ and $n_i$ is the impurity density.

To qualitatively discuss the spin-exchange field \eqref{elf} and spin relaxation time  [Eqs.~(\ref{LongiSRT}) and (\ref{TransSRT})],
we consider two regimes. i) $\beta\epsilon_L\gg 1$, i.e., the strongly magnetized regime with large $B$ and low $T$, the spins of local moments are fixed and the flip of itinerant electrons' spins is prohibited. Thus, the spin-exchange field~\eqref{elf} is strongest, i.e.,  $\mathcal{B}_{sd}=-n_{
\mathrm{S}}\mathcal{J}S/(\hbar \textsl{g}\mu_B)$, while the spin relaxation time [Eqs.~(\ref{LongiSRT}) and (\ref{TransSRT})] becomes anisotropic, i.e.,  $\tau^{-1}_{\Vert}=\tau^{-1}_{0}$ and $\tau^{-1}_{\perp}=\tau^{-1}_{0}+   \Omega_{0}$ with $\Omega_{0}=\frac{\pi}{\hbar}   n_{\text{S}}  \nu_F (\mathcal{J} S
)^2$, as a result of the vanishing spin-flip rate and the maximum spin dephasing rate. ii) In the opposite limit ($\beta\epsilon_L\ll 1$) with small $B$ and high $T$, where the microscopic ferromagnetic order is destroyed, the spin-exchange field~\eqref{elf} vanishes because $\langle S_{\Vert
}\rangle=0$, while the spin relaxation time [Eqs.~(\ref{LongiSRT}) and (\ref{TransSRT})] become isotropic, i.e.,    $\tau^{-1}_{\Vert}=\tau^{-1}_{\perp}=\tau^{-1}_{0}+   \Omega_{1}$, where $\Omega_{1}=\frac{2\pi}{3\hbar}\nu_Fn_{\text{S}}\mathcal{J}^2S(S+1)$. Consequently, the spin relaxation time and spin-exchange field induced by the spin-exchange coupling strongly rely on $B$ and $T$, and below we will see they are crucial for the $B$- and $T$-dependent MR.

Provided by boundary
conditions~\eqref{mfvavkdfmv}, the diffusion equation (\ref{3.3}) can be
analytically solved and we obtain the widespread phenomenological formula of the longitudinal and transverse resistivity (see derivations in Appendix~\ref{resistivity})
\begin{equation}  \label{SL}
\Delta\rho _{\mathrm{L}}\simeq 2\theta _{\mathrm{SH}}^{2}\rho _{\mathrm{L0}}-\theta _{\mathrm{SH}}^{2}\Delta
\rho _{0}+\theta _{\mathrm{SH}}^{2}\Delta \rho _{1}\left( 1-\hat{m}%
_{y}^{2}\right) ,
\end{equation}%
\begin{equation}
\rho _{T}=\theta _{\mathrm{SH}%
}^{2}\Delta \rho _{1}\hat{m}%
_{x}\hat{m}_{y}-\theta _{\mathrm{SH}}^{2}\Delta \rho _{2}
\hat{m}_{z},
\end{equation}
with 
\begin{equation} \label{vvh0}
\frac{\Delta \rho _{0}}{\rho _{\mathrm{L0}}}=\frac{2\ell _{\Vert }}{d_{N}}\tanh \left( \frac{%
d_{N}}{2\ell _{\Vert }}\right),
\end{equation}%
\begin{equation} \label{vvh1}
\frac{\Delta \rho _{1}}{\rho _{\mathrm{L0}}}=\frac{2\ell _{\Vert  }}{d_{N}}\tanh \left( \frac{%
d_{N}}{2\ell _{\Vert  }}\right)-\mathrm{Re}%
\left[ \frac{2\Lambda}{d_{N}}\tanh \left( \frac{%
d_{N}}{2\Lambda}\right)\right],
\end{equation}%
\begin{equation} \label{vvh2}
\frac{\Delta \rho _{2}}{\rho _{\mathrm{L0}}}=\mathrm{Im}%
\left[ \frac{2\Lambda}{d_{N}}\tanh \left( \frac{%
d_{N}}{2\Lambda}\right)\right],
\end{equation}%
where $\rho _{\mathrm{L0}}=1/\sigma_{\text{D}}$
and 
$
\Lambda^{-2}=\ell _{\mathrm{\bot }}^{-2}+i\ell _{\mathrm{L}}^{-2}$. Here, the order parameter is the magnetization of the ferromagnet. Equations (\ref{SL}-\ref{vvh2}), as the central result, not only microscopically explain the universal cosine-square law of anisotropic MR with the magnetization direction ($\hat{m}_y\equiv\sin\alpha$)  but also quantitatively describe $B$- and $T$-dependent MR. Importantly, both $T $ and $B$ dependencies of $\Delta
\rho _{0}$ and $\Delta
\rho _{1}$, i.e., Eqs.~\eqref{vvh0} and \eqref{vvh1}, are
reflected in Hanle spin precession length $\ell _{\mathrm{L}}$, spin diffusion lengths $\ell _{\mathrm{\bot }} $ and $\ell _{\mathrm{\parallel }}$ through the spin-exchange field \eqref{elf}, spin relaxation time \eqref{LongiSRT} and \eqref{TransSRT}, respectively. The dependence of resistance on $T$ arises from $\langle S_{\Vert}\rangle $ and $\langle S_{\Vert}^{2}\rangle $, while that of $B$ has an extra
channel - the magnetic-field spin precession frequency $
\omega _{B}(\propto B)$.

Then we investigate the microscopic mechanisms of various MR effects using our formulas (\ref{SL}-\ref{vvh2}). Note that the spin-exchange coupling  in magnetic materials is ubiquitous and profoundly affects the MR effects. At $\mathcal{J}_{sd}=0$, the spin diffusion lengths become isotropic (i.e.,  $\ell_{\Vert,\perp}=\ell_0=\sqrt{\mathcal{D}\tau_0}$) and we recover the previous theory of spin precession MR (i.e., Hanle MR~\cite{velez2016hanle}) where $\Delta \rho_1\propto B^2$ for small $B$. Next, we focus on the case of $\mathcal{J}_{sd} \neq 0$. i) The spin-exchange coupling shifts the Hanle spin precession frequency $\omega_L=g\mu_B(B-\mathcal{B}_{sd})$, introduces a finite value of $\Delta \rho _{1}\propto (B-\mathcal{B}_{sd})^2$ even when the anisotropic spin relaxation is artificially removed by setting $\ell_{\Vert}= \ell_{\perp}$, and contributes to a shifted spin precession MR. Thus, our theory, different from the previous spin precession MR independent of $T$,  effectively includes $T$ and $B$ dependencies of MR through the spin-exchange field \eqref{elf}. ii) The spin-exchange coupling causes anisotropic spin diffusion lengths ($\ell_{\Vert}\neq \ell_{\perp}$), produces a finite value of $\Delta \rho _{1}=\frac{2\ell _{\Vert  }}{d_{N}}\tanh \left( \frac{%
d_{N}}{2\ell _{\Vert  }}\right)-\frac{2\ell _{\perp  }}{d_{N}}\tanh \left( \frac{%
d_{N}}{2\ell _{\perp }}\right)$ when artificially setting $\omega_L=0$, and  accounts for anisotropic spin relaxation MR. Thus, our microscopic theory, exceeding the previous phenomenological theory of anisotropic MR, adequately captures the $T$ and $B$ dependencies of $\Delta\rho_{1}$ through the spin relaxation time \eqref{LongiSRT} and \eqref{TransSRT}. iii) The $B$ modulation of $\Delta\rho _{0}$, i.e., Eq.~\eqref{vvh0}, depends only on the longitudinal spin relaxation time~\eqref{LongiSRT} originating entirely
from spin-flip processes and this MR is unambiguously associated with magnon emission and absorption, thus referring to isotropic magnon MR. Overall, we demonstrate the isotropic magnon, anisotropic spin relaxation and spin precession MR originating from the magnon-induced spin flip, anisotropic spin relaxation, and Hanle spin precession of itinerant electrons, respectively.

\subsection{Antiferromagnetic MR} \label{antiferromagneticMR}

Next, we  work on the  antiferromagnetic MR with two sublattices. For simplicity, we merely consider zero external magnetic field, where the spin-exchange fields from two sublattices cancel with each and there is not Zeeman magnetic field anymore (i.e., $\omega_L=0$).  The key role of the spin-exchange coupling, Eq.~\eqref{spin-exchange}, in the antiferromagnetic MR effect lies in its induction of anisotropic spin relaxation for free electrons. Assuming isotropic spin-exchange coupling again ($\mathcal{J}^{\nu}_{\imath}=\mathcal{J}$), the corresponding longitudinal and transverse spin relaxation times, [i.e., Eqs.~\eqref{mLTdppaAdU} and \eqref{mfjanwegn}]  are given, respectively, by
\begin{align} \label{aLongiSRT}
	\frac{1}{\tau_{\Vert}} &= \frac{1}{\tau_0}
	+\frac{\pi}{\hbar} n^{\imath}_{\text{S}}\nu_F\mathcal{J}^2 \beta\epsilon_L^{\imath}
	n_{B}(\epsilon^{\imath}_L)[\text{S}(\text{S}+1)-\langle \text{S}_{\imath}^{\Vert}\text{S}_{\imath}^{\Vert}\rangle - \langle \text{S}_{\imath}^{\parallel}\rangle],
\end{align}
\begin{align}  \label{aTransSRT}
	\frac{1}{\tau_{\perp}} &= \frac{1}{2\tau_0} + \frac{1}{2\tau_{\Vert}}
	+ \frac{\pi}{\hbar} n^{\imath}_{\text{S}} \nu_F \mathcal{J}^2 \langle \text{S}_{\imath}^{\Vert} \text{S}_{\imath}^{\Vert} \rangle,
\end{align}
where $n^{\imath}_{\mathrm{S}}$ is the density of $\imath$-sublattice local moments. Again, the second term in Eq.~\eqref{aLongiSRT} corresponds to the spin-flip rate arising from magnon emission and absorption in antiferromagnetic configuration, where $\mathrm{S}_{\imath}^{\Vert}$ denotes the spin component parallel to the Néel field direction. Besides, the third term in Eq.~\eqref{aTransSRT} accounts for the spin dephasing rate arising from scattering processes, during which the electron’s spin conserves but accumulates a precession phase~\cite{zhang2019theory}. The anisotropic spin relaxation time induced by the spin-exchange coupling depends strongly on $T$ through $\langle \text{S}^{\Vert}_{\imath} \rangle$ and $\langle \text{S}^{\parallel}_{\imath} \text{S}^{\parallel}_{\imath} \rangle$, and is consequently responsible for the $T$-dependent behavior of MR effect in antiferromagnet.

With the boundary condition [Eq.\eqref{mfvavkdfmv}], the anisotropic spin diffusion equation [Eq.\eqref{3.3}] can be analytically solved~\cite{zhang2024altermagnet}. The resulting expression for the longitudinal resistivity reads
\begin{align} \label{aSL}
\Delta\rho _{\mathrm{L}}&\simeq \theta _{\mathrm{SH}%
}^{2}\Delta\rho _{1} \left( 1-n_{y}^{2}\right)-\theta _{\mathrm{SH}}^{2}\Delta\rho _{0},
\end{align}%
\begin{equation}
\rho _{T}=\theta _{\mathrm{SH}%
}^{2}\Delta \rho _{1}\hat{n}%
_{x}\hat{n}_{y},
\end{equation}
with
\begin{equation} \label{avvh0}
\frac{\Delta\rho _{0}}{\rho _{\mathrm{L0}}}=\frac{2\ell _{\Vert }}{d_{N}}\tanh \left( \frac{%
d_{N}}{2\ell _{\Vert }}\right),
\end{equation}%
\begin{equation} \label{avvh}
\frac{\Delta\rho _{1}}{\rho _{\mathrm{L0}}}=\frac{2\ell _{\Vert  }}{d_{N}}\tanh \left( \frac{%
d_{N}}{2\ell _{\Vert  }}\right)- \frac{2\ell _{\perp  }}{d_{N}}\tanh \left( \frac{%
d_{N}}{2\ell _{\perp  }}\right),
\end{equation}%
\begin{equation} \label{avvh1}
\frac{\Delta \rho _{2}}{\rho _{\mathrm{L0}}}=\mathrm{Im}%
\left[ \frac{2\ell _{\perp  }}{d_{N}}\tanh \left( \frac{%
d_{N}}{2\ell _{\perp  }}\right)\right].
\end{equation}%
Here, the order parameter is the Néel vector rather the magnetization in ferromagnetic magnetoresistance. Equations~\eqref{aSL}--\eqref{avvh1} represent the central results of the antiferromagnetic MR. They reveal a novel anisotropy in the MR effect that depends on the angle between the Néel vector and $y$ axis ($n_y=\sin\alpha$) and provide microscopic insight into the MR mechanism via two distinct contributions: the magnon-induced spin-flip process [Eq.~\eqref{avvh0}] and the anisotropic spin relaxation of free electrons [Eq.~\eqref{avvh}].

In this case, the MR effect follows a cosine-square dependence,
$\theta_{\mathrm{SH}}^{2} \Delta\rho_1 (1 - n_y^2) = \theta_{\mathrm{SH}}^{2} \Delta\rho_1 \cos^2\alpha$, which exhibits symmetric behavior with respect to $+n_y$ and $-n_y$, and displays a $\pi$-periodic oscillation. The resulting $\pi$-periodic MR with respect to the Néel vector—rather than the net magnetization—resembles the anisotropic MR typically observed in antiferromagnets. This MR anisotropy originates from anisotropic spin relaxation, specifically the difference between $\ell_{\parallel}$ and $\ell_{\perp}$ [see Eq.\eqref{vvh1}]. To provide an intuitive explanation, we focus on the altermagnetic phase at sufficiently low temperature ($T \ll T_N$). When the spin polarization of free electrons is aligned with the Néel vector, spin relaxation is weak, such that $\tau^{-1}_{\parallel} \rightarrow \tau^{-1}_0$. As a result, more transverse spin current is reflected at the top and bottom boundaries and subsequently converted into a longitudinal charge current via the inverse spin Hall effect, leading to a low-resistance state.  In contrast, when the spin polarization is perpendicular to the Néel vector, the spin relaxation becomes significantly stronger, i.e., $\tau^{-1}_{\perp} = \tau^{-1}_0 + 2\Omega_0$, where $\Omega_0 = \frac{\pi}{\hbar} n_s \nu_F \mathcal{J}^2 \text{S}^2$. In this regime, the transverse spin current is more likely to decay during propagation and thus fails to be converted into a charge current via the inverse spin Hall effect, resulting in a high-resistance state.

\section{Summary}
We have developed a comprehensive microscopic theory of MR from an open-quantum-system perspective. In particular, we solve the Liouville–von Neumann equation for a hybrid system of itinerant electrons and local moments using the time-convolutionless projection-operator technique.  Both ferromagnetic and antiferromagnetic MR are described in terms of temperature- and field-dependent spin decoherence processes (spin relaxation and spin dephasing), where the resistance varies with the magnetization and the Néel vector, respectively. Our theory provides a deeper understanding of the fundamental mechanisms underlying MR and offers insights relevant to experiments on magnetic materials. Furthermore, the framework can be readily extended to a broader class of magnetic systems, including ferrimagnetic and antiferromagnetic metals and semiconductors, as well as normal metals decorated with magnetic impurities (see Appendix~\ref{abgvkfk}), and furthermore our quantum spin decoherence theory can be applied to macroscopic magnetic materials~\cite{zhang2025anomalous,zhang2025theory}.

\section{Acknowledgment} 

This work is supported by National Key R$\&$D Program of China (Grant No. 2020YFA0308800, No. 2022YFA1402600, and No. 2022YFA1403800), the National Natural Science Foundation of China (Grant No. 12234003, No. 12321004, No. 12274027, and No. 12374122),  and the Hong Kong Research Grants Council Grants (No. 16300523), and the Guangdong Basic and Applied Basic Research Foundation (Grants No. 2024A1515030118).

\appendix

\setcounter{section}{0}
\setcounter{equation}{0}

\section{Derivations of Eq.~\eqref{mfagagttgt1}} \label{fvadkfvk}

In this section, we treat magnetoresistance effect as an open-quantum problem and derive collision integral within second-order Born-Markov approximation and Weiss mean-field approximation [Eq.~\eqref{mfagagttgt1} in main text].

The quantum kinetic equation of itinerant electrons is the time evolution of one-electron density matrix, defined as
\begin{equation} \label{jgsufgnn}
    \varrho_{\alpha_2\alpha_1}(t)= \langle \alpha_2\vert \hat{\varrho}(t) \vert \alpha_1\rangle=\mathrm{Tr}_e\{\hat{f}^+_{\alpha_1}\hat{f}^{}_{\alpha_2}\hat{\rho}(t)\}.
\end{equation}
The time evolution of each element of density matrix is 
\begin{equation} \label{RFE} 
\dfrac{\partial}{\partial t} \varrho_{\alpha_2\alpha_1}(t)-\frac{i}{\hbar}[\varrho(t), \hat{\mathcal{H}}^e_0]_{\alpha_2\alpha_1}-\frac{i}{\hbar}[\varrho(t), \hat{\mathcal{V}}_m]_{\alpha_2\alpha_1}=\mathcal{ J}_{\alpha_2\alpha_1}(\hat{\varrho}\vert t),
\end{equation} 
with
\begin{eqnarray} 
  \hat{\mathcal{V}}_m =-\frac{1}{2} \sum_{\jmath \nu} N_{\jmath}  \mathcal{J}^{\nu}_{\jmath}   \langle \widehat{S}^{\nu}_{\jmath}\rangle \hat{\sigma}^{\nu}  ,
\end{eqnarray}
where $\alpha_i=(\vec{p}_i,s_i)$. The collision integral in the right hand side of Eq. (\ref{RFE}) is given by 
\begin{align} \label{CIRF}
 \mathcal{ J}_{\alpha_2\alpha_1}(\hat{\varrho}\vert t)&=\sum_{n_i}  \int_{0}^{t-t_0} d\tau  \mathrm{Tr}_e\left\lbrace   \left[ \hat{V}^{n_1},\hat{f}^+_{\alpha_1}\hat{f}^{}_{\alpha_2} \right]\right.\\
 &\times \left. \left[\mathcal{D}_{n_1n_2}(+\tau)  \hat{\mathcal{V}}^{n_2}(-\tau) \hat{\rho}(t)\right.\right.\notag \\
 &-\left.\left.\mathcal{D}_{n_2n_1}(-\tau)\hat{\rho}(t)\hat{\mathcal{V}}^{n_2}(-\tau)\right]  \right\rbrace. \notag
\end{align}
It is required to calculate commutator 
\begin{align} \label{fhanfu}
    [V^{n_1},\hat{f}^+_{\alpha_1}\hat{f}^{}_{\alpha_2}]
    &=\sum_{\alpha_i}  \left[\langle \alpha_3\vert \hat{V}^{n_1} \vert \alpha_1\rangle \hat{f}^{+}_{\alpha_3} \hat{f}^{}_{\alpha_2} \right.\\
    &-\left.\langle \alpha_2\vert \hat{V}^{n_1} \vert \alpha_3\rangle \hat{f}^{+}_{\alpha_1} \hat{f}^{}_{\alpha_3}\right].\notag 
\end{align}
Substituting Eq. \eqref{fhanfu} into \eqref{CIRF}, we reach
\begin{widetext}
\begin{align} \label{CIRFfanjgra}
\mathcal{ J}_{\alpha_2\alpha_1}(\hat{\varrho}\vert t)&= \sum_{n_i\alpha_i}  \int_{0}^{t-t_0} d\tau  \mathrm{Tr}_e\left\lbrace   \left[\langle \alpha_3\vert \hat{V}^{n_1} \vert \alpha_1\rangle \hat{f}^{+}_{\alpha_3} \hat{f}^{}_{\alpha_2}-\langle \alpha_2\vert \hat{V}^{n_1} \vert \alpha_3\rangle \hat{f}^{+}_{\alpha_1} \hat{f}^{}_{\alpha_3}\right]\right.\\
 &\times \left. \left[\mathcal{D}_{n_1n_2}(+\tau)  \hat{\mathcal{V}}^{n_2}(-\tau) \hat{\rho}(t)-\mathcal{D}_{n_2n_1}(-\tau)\hat{\rho}(t)\hat{\mathcal{V}}^{n_2}(-\tau)\right]  \right\rbrace. \notag
\end{align}   
\end{widetext}
By substitution of 
\begin{equation}
 \hat{\mathcal{V}}^{n_2}(-\tau)=\sum_{\alpha_i} e^{+i\omega^{\alpha_6}_{\alpha_5}\tau}\langle \alpha_5\vert \hat{V}^{n_2} \vert \alpha_6\rangle \hat{f}^+_{\alpha_5}\hat{f}^{}_{\alpha_6},
\end{equation} 
Eq. \eqref{CIRFfanjgra} becomes 
\begin{widetext}
\begin{align} \label{fagayotar1}
 \mathcal{ J}_{\alpha_2\alpha_1}(\hat{\varrho})&=  \sum_{n_i\alpha_i} \mathrm{Tr}_e\left\lbrace   \left[\langle \alpha_3\vert \hat{V}^{n_1} \vert \alpha_1\rangle \hat{f}^{+}_{\alpha_3} \hat{f}^{}_{\alpha_2}-\langle \alpha_2\vert \hat{V}^{n_1} \vert \alpha_3\rangle \hat{f}^{+}_{\alpha_1} \hat{f}^{}_{\alpha_3}\right]\right.\\
 &\times \left. \left[\mathcal{D}^+_{n_1n_2}(\omega^{\alpha_6}_{\alpha_5})  \langle \alpha_5\vert \hat{V}^{n_2} \vert \alpha_6\rangle \hat{f}^+_{\alpha_5}\hat{f}^{}_{\alpha_6} \hat{\rho}(t)-\mathcal{D}^-_{n_2n_1}(\omega^{\alpha_5}_{\alpha_6})\hat{\rho}(t)\langle \alpha_5\vert \hat{V}^{n_2} \vert \alpha_6\rangle \hat{f}^+_{\alpha_5}\hat{f}^{}_{\alpha_6}\right]  \right\rbrace, \notag
\end{align}    
\end{widetext}
with
\begin{align} \label{fdkaknp}
    \mathcal{D}^{+}_{n_1n_2}(\omega) =\int^{+\infty}_{0} d\tau e^{+i\omega\tau-\eta \tau}\mathcal{D}_{n_1n_2}(\tau),
\end{align}
\begin{align} \label{fdkaknn}
    \mathcal{D}^{-}_{n_1n_2}(\omega) =\int^{0}_{-\infty} d\tau e^{+i\omega\tau+\eta \tau}\mathcal{D}_{n_1n_2}(\tau),
\end{align}
where $t_0\rightarrow -\infty$, and  $\eta\rightarrow 0^+$ was introduced to remove the divergence of the infinity integral.  
Eq. \eqref{fagayotar1} can be divided into two terms 
\begin{align} \label{CIRFfanjgr}
\hat{\mathcal{J}}_{\alpha_2\alpha_1}(\varrho)&=\hat{\mathcal{J}}^{e}_{\alpha_2\alpha_1}(\hat{\varrho})+\hat{\mathcal{J}}^{a}_{\alpha_2\alpha_1}(\hat{\varrho}),
\end{align}
with 
\begin{align} \label{vjwpa1}
   \hat{\mathcal{J}}^{e}_{\alpha_2\alpha_1}(\hat{\varrho})&= \sum_{n_i\alpha_i} \sum_{n_i\alpha_i}\mathcal{D}^-_{n_2n_1}(\omega^{\omega_5}_{\alpha_6})\langle \alpha_5\vert \hat{V}^{n_2} \vert \alpha_6\rangle\\
   &\times\left [  \langle \alpha_2\vert \hat{V}^{n_1} \vert \alpha_3\rangle R^{\alpha_5\alpha_1}_{\alpha_6\alpha_3}(t)-\langle \alpha_3\vert \hat{V}^{n_1} \vert \alpha_1\rangle  R^{\alpha_5\alpha_3}_{\alpha_6\alpha_2}(t) \right],\notag 
\end{align}
\begin{align} \label{vjwpa2}
   \hat{\mathcal{J}}^{a}_{\alpha_2\alpha_1}(\hat{\varrho})&= \sum_{n_i\alpha_i} \sum_{n_i\alpha_i}\mathcal{D}^+_{n_1n_2}(\omega^{\omega_6}_{\alpha_5})\langle \alpha_5\vert \hat{V}^{n_2} \vert \alpha_6\rangle \\
   &\times \left[\langle \alpha_3\vert \hat{V}^{n_1} \vert \alpha_1\rangle  R^{\alpha_3\alpha_5}_{\alpha_2\alpha_6}(t)-\langle \alpha_2\vert \hat{V}^{n_1}\vert \alpha_3\rangle R^{\alpha_1\alpha_5}_{\alpha_3\alpha_6}(t) \right].\notag 
\end{align}
The many-electron correlation functions, are defined as
\begin{align} \label{lbneruyt}
    R^{\alpha_1\alpha_3}_{\alpha_2\alpha_4}(t)=\text{Tr}_e\left\{\hat{f}^+_{\alpha_1}\hat{f}^{}_{\alpha_2}\hat{f}^{+}_{\alpha_3} \hat{f}^{}_{\alpha_4}\hat{\rho}(t)\right\},
\end{align}
To calculate Eqs. \eqref{vjwpa1} and \eqref{vjwpa2}, it is required to calculate  many-electron correlation function \eqref{lbneruyt}. To reach a normal order inside the trace $\text{Tr}_e$, we reach 
\begin{align} \label{kkvpgbw1}
    R^{\alpha_1\alpha_3}_{\alpha_2\alpha_4}(t)=\delta_{\alpha_2\alpha_3}\varrho_{\alpha_4\alpha_1}-\text{Tr}_e\left\{\hat{f}^+_{\alpha_1}\hat{f}^{+}_{\alpha_3}\hat{f}^{}_{\alpha_2} \hat{f}^{}_{\alpha_4}\hat{\rho}(t)\right\}.
\end{align} 
The second term in normal order, can be approximated into 
\begin{align} \label{kkvpgbw2}
    \text{Tr}_e\left\{\hat{f}^+_{\alpha_1}\hat{f}^{+}_{\alpha_3}\hat{f}^{}_{\alpha_2} \hat{f}^{}_{\alpha_4}\check{\rho}_e(t)\right\}\simeq \varrho_{\alpha_4\alpha_1}\varrho_{\alpha_2\alpha_3}-\varrho_{\alpha_4\alpha_3}\varrho_{\alpha_2\alpha_1}.
\end{align}
Hereafter, we omit the disconnected terms, such as $\varrho_{\alpha_4\alpha_3}\varrho_{\alpha_2\alpha_1}$. 
Then, we attain 
\begin{align} \label{kkdvpgbw1}
    R^{\alpha_1\alpha_3}_{\alpha_2\alpha_4}(t)\simeq \bar{\varrho}_{\alpha_2\alpha_3}\varrho_{\alpha_4\alpha_1},
\end{align}
with 
\begin{align}
    \bar{\varrho}_{\alpha_1\alpha_2}=\delta_{\alpha_1\alpha_2}-\varrho_{\alpha_1\alpha_2}.
\end{align}
With the help from  Eqa. \eqref{kkvpgbw1} and \eqref{kkvpgbw2}, let us calculate term by term
\begin{widetext}
\begin{align} \label{farnyay1}
   \hat{\mathcal{J}}^{e}_{\alpha_2\alpha_1}(\hat{\varrho})
   &= \sum_{n_i\alpha_i}  \mathcal{D}^-_{n_2n_1}(\omega^{\omega_5}_{\alpha_6})[\langle \alpha_2\vert \hat{V}^{n_1} \vert \alpha_3\rangle\varrho_{\alpha_3\alpha_5}\langle \alpha_5\vert \hat{V}^{n_2} \vert \alpha_6\rangle \bar{\varrho}_{\alpha_6\alpha_1}-\varrho_{\alpha_2\alpha_5}\langle \alpha_5\vert \hat{V}^{n_2} \vert \alpha_6\rangle\bar{\varrho}_{\alpha_6\alpha_3}\langle \alpha_3\vert \hat{V}^{n_1} \vert \alpha_1\rangle ],
\end{align}
\begin{align}  \label{farnyay4}
   \hat{\mathcal{J}}^{a}_{\alpha_2\alpha_1}(\hat{\varrho})
   &= \sum_{n_i\alpha_i}  \mathcal{D}^+_{n_1n_2}(\omega^{\omega_6}_{\alpha_5}) [\bar{\varrho}_{\alpha_2\alpha_5} \langle \alpha_5\vert \hat{V}^{n_2} \vert \alpha_6\rangle \varrho_{\alpha_6\alpha_3}  \langle \alpha_3\vert \hat{V}^{n_1} \vert \alpha_1\rangle
   -\langle \alpha_2\vert \hat{V}^{n_1} \vert \alpha_3\rangle \bar{\varrho}_{\alpha_3\alpha_5}\langle \alpha_5\vert \hat{V}^{n_2} \vert \alpha_6\rangle\varrho_{\alpha_6\alpha_1} ].
\end{align}    
\end{widetext}
Here we are interested in the third line of collision terms \eqref{farnyay1}-\eqref{farnyay4}. 
For the sake of simplicity, we assume the plane wave function
\begin{align}
   \langle \vec{r} \vert\alpha\rangle=\frac{1}{\sqrt{V}} e^{\vec{k}\cdot\vec{r}} \vert\sigma\rangle.
\end{align}
Thus, we reach  
\begin{widetext}
\begin{align} \label{fagagttgt1}
   \mathcal{J}^{\sigma_2\sigma_1}_{\vec{k}_2\vec{k}_1}(\check{\varrho})&=\frac{1}{V^2} \sum_{n_i\alpha_i} \left\lbrace \mathcal{D}^-_{n_2n_1}(\omega^{\omega_5}_{\alpha_6})\left[e^{-i(\vec{k}_2-\vec{k}_3)\cdot \vec{r}_{j_1}-i(\vec{k}_5-\vec{k}_6)\cdot \vec{r}_{j_2}} \sigma_{\sigma_2\sigma_3}^{\nu_1} \varrho^{\sigma_3\sigma_5}_{\vec{k}_3\vec{k}_5} \sigma_{\sigma_5\sigma_6}^{\nu_2}  \bar{\varrho}^{\sigma_6\sigma_1}_{\vec{k}_6\vec{k}_1}\right.\right.\\
    &-\left.e^{-i(\vec{k}_5-\vec{k}_6)\cdot \vec{r}_{j_2}-i(\vec{k}_3-\vec{k}_1)\cdot \vec{r}_{j_1}} \varrho^{\sigma_2\sigma_5}_{\vec{k}_2\vec{k}_5} \sigma_{\sigma_5\sigma_6}^{\nu_2}   \bar{\varrho}^{\sigma_6\sigma_3}_{\vec{k}_6\vec{k}_3} \sigma_{\sigma_3\sigma_1}^{\nu_1}\right] \notag \\
  &+\mathcal{D}^+_{n_1n_2}(\omega^{\omega_6}_{\alpha_5})\left[e^{-i(\vec{k}_5-\vec{k}_6)\cdot \vec{r}_{j_2}-i(\vec{k}_3-\vec{k}_1)\cdot \vec{r}_{j_1}} \bar{\varrho}^{\sigma_2\sigma_5}_{\vec{k}_2\vec{k}_5}  \sigma_{\sigma_5\sigma_6}^{\nu_2}  \varrho^{\sigma_6\sigma_3}_{\vec{k}_6\vec{k}_3}   \sigma_{\sigma_3\sigma_1}^{\nu_1}\right.  \notag \\
  &-\left.\left. e^{-i(\vec{k}_2-\vec{k}_3)\cdot \vec{r}_{j_1}-i(\vec{k}_5-\vec{k}_6)\cdot \vec{r}_{j_2}}  \sigma_{\sigma_2\sigma_3}^{\nu_1}  \bar{\varrho}^{\sigma_3\sigma_5}_{\vec{k}_3\vec{k}_5} \sigma_{\sigma_5\sigma_6}^{\nu_2}    
   \varrho^{\sigma_6\sigma_1}_{\vec{k}_6\vec{k}_1}\right]\right\}.\notag 
\end{align}    
\end{widetext}
The spin-spin correlation functions of quantum bath in Eq. \eqref{fagagttgt1} are depended on the magnetic configuration of magnetic moments.

\section{Derivations of Eqs. (\ref{SL}-\ref{vvh2})} \label{resistivity}

In this section, we  present the detailed derivations of longitudinal resistivity [Eqs. (\ref{SL}-\ref{vvh2}) in main text] by solving the spin diffusion equation [Eq.~\eqref{3.3} in main text] with boundary condition [Eq.~\eqref{mfvavkdfmv}].

Let's start with the diffusion equation for spin density in the presence of the Hanle spin precession. Our starting point is the spin diffusion  equation \eqref{3.3} given as follows  
\begin{equation}
\nabla ^{2}\mu _{\mathrm{s}i }=\mathcal{M}_{ij}\mu _{%
\mathrm{s}j }=\left( \mathcal{M}_{ij}^{\mathrm{\bot }}+\mathcal{M}%
_{ij}^{\mathrm{m}}+\mathcal{M}_{ij}^{\mathrm{L}}\right) \mu _{\mathrm{s}%
\jmath },  \label{3.2}
\end{equation}%
with
\begin{align}
    \mathcal{M}_{ij}^{\mathrm{\bot }}=\delta _{ij}\ell _{\mathrm{\bot }}^{-2},
\end{align}
\begin{align}
    \mathcal{M}_{ij}^{\mathrm{m}}=\hat{m}_{i}\hat{m}_{j}(\ell _{\Vert}^{-2}-\ell _{\mathrm{\bot }}^{-2}),
\end{align}
\begin{equation}
\mathcal{M}_{ij}^{\mathrm{L}}=-\ell _{\mathrm{L}}^{-2}\epsilon _{ijk}m_{k}.
\end{equation}%
where $\ell _{\mathrm{%
L}}=\sqrt{\mathcal{D}/\tilde{\omega}_{L}}$ is spin-precession lengths, while $\ell _{\mathrm{\bot }}=\sqrt{\mathcal{D}\tau _{\mathrm{\bot }}}$ and $%
\ell _{\Vert}=\sqrt{\mathcal{D}\tau _{\Vert}}$  are transverse and longitudinal spin-diffusion length, respectively, and $\mathcal{D}%
=\sigma _{\mathrm{D}}/\left( 2\nu_F e^{2}\right) $ is diffusion coefficent.
There are two kinds of coupling in diffusion equation (\ref{3.2}). They are $%
\mathcal{M}^{\mathrm{m}}$ terms and $\mathcal{M}^{\mathrm{L}}$ terms. The
significant difference of these two couplings is that the $\mathcal{M}^{%
\mathrm{m}}$ is symmetric, while $\mathcal{M}^{\mathrm{L}}$ is
antisymmetric. The electrons are diffused by eigen-modes of diffusion matrix
$\mathcal{M}$, defined as follows, 
\begin{align}
    \ell _{\nu }^{-2}\mathcal{E}_{i\nu }=%
\mathcal{M}_{ij}\mathcal{E}_{j\nu }.
\end{align}
The eigenvalues of
diffusion matrix $\mathcal{M}$ are given by 
\begin{equation}
\ell _{1}^{-2}=\ell _{\mathrm{\bot }}^{-2}+i\ell _{\mathrm{L}}^{-2},\ell
_{2}^{-2}=\ell _{\mathrm{\bot }}^{-2}-i\ell _{\mathrm{L}}^{-2},\ell
_{3}^{-2}=\ell _{\mathrm{\bot }}^{-2}+\ell _{\mathrm{m}}^{-2}=\ell _{\mathrm{%
\parallel }}^{-2},
\end{equation}%
while  the  eigenvectors of
diffusion matrix $\mathcal{M}$ are given by
\begin{equation}
\mathcal{E}=%
\begin{bmatrix}
\frac{\hat{m}_{1}^{2}-1}{\sqrt{2}N_{r}} & \frac{\hat{m}_{1}^{2}-1}{\sqrt{2}%
N_{r}} & \hat{m}_{1} \\
\frac{+i\hat{m}_{3}+\hat{m}_{1}\hat{m}_{2}}{\sqrt{2}N_{r}} & \frac{-i\hat{m}%
_{3}+\hat{m}_{1}\hat{m}_{2}}{\sqrt{2}N_{r}} & \hat{m}_{2} \\
\frac{-i\hat{m}_{2}+\hat{m}_{1}\hat{m}_{3}}{\sqrt{2}N_{r}} & \frac{+i\hat{m}%
_{2}+\hat{m}_{1}\hat{m}_{3}}{\sqrt{2}N_{r}} & \hat{m}_{3}%
\end{bmatrix}%
,  \label{EV}
\end{equation}%
\begin{equation}
\mathcal{E}^{-1}=%
\begin{bmatrix}
\frac{\hat{m}_{1}^{2}-1}{\sqrt{2}N_{r}} & \frac{-i\hat{m}_{3}+\hat{m}_{1}%
\hat{m}_{2}}{\sqrt{2}N_{r}} & \frac{+i\hat{m}_{2}+\hat{m}_{1}\hat{m}_{3}}{%
\sqrt{2}N_{r}} \\
\frac{\hat{m}_{1}^{2}-1}{\sqrt{2}N_{r}} & \frac{+i\hat{m}_{3}+\hat{m}_{1}%
\hat{m}_{2}}{\sqrt{2}N_{r}} & \frac{-i\hat{m}_{2}+\hat{m}_{1}\hat{m}_{3}}{%
\sqrt{2}N_{r}} \\
\hat{m}_{1} & \hat{m}_{2} & \hat{m}_{3}%
\end{bmatrix}%
,  \label{IEV}
\end{equation}%
with $N_{r}=\sqrt{(1-\hat{m}_{1}^{2})}$. Obviously, $\left[ \mathcal{M}^{%
\mathrm{m}},\mathcal{M}^{\mathrm{L}}\right] =0$ $\left( \left[ \mathcal{M}^{%
\mathrm{\bot }},\mathcal{M}^{\mathrm{L}}\right] =0\right) $, $\mathcal{E}%
_{i\nu }$ are independent of spin-diffusion length $\ell _{\Vert/\perp}^{-2}$
(spin-precession lengths $\ell _{\mathrm{L}}^{-2}$).

For simplicity, we consider a system homogeneous in the $x$-$y$ plane,
we focus on the spin current density flowing in the $\hat{z}$-direction. The boundary conditions require that no spin current can enter or exit
the 2D conductive channel in $\hat{z}$-direction. Thus, we obtain boundary conditions at top ($z=0$) and bottom ($z=d_N$)  interfaces~\cite{chen2013theory,velez2016hanle} 
\begin{equation}
\vec{0}=-\frac{\sigma _{\mathrm{D}}}{2e}\left. \partial_z \vec{\mu}_{\mathrm{s}%
}(z)\right\vert _{z=0}-J_{\mathrm{SH}}\hat{y},  \label{BC.1}
\end{equation}%
\begin{equation}
\vec{0}=-\frac{\sigma _{\mathrm{D}}}{2e}\left. \partial_z \vec{\mu}_{\mathrm{s}%
}(z)\right\vert _{z=d_N}-J_{\mathrm{SH}}\hat{y},  \label{BC.11}
\end{equation}%
where $J_{\mathrm{SH}}=\theta _{\mathrm{SH}}\sigma _{D}E_{x}$ is the
bare spin Hall current, i.e., the spin current generated directly by the
spin Hall effect. 

The solution of diffusion equation (\ref{3.2}) can be assume that
\begin{equation}
\mu _{\mathrm{s}i}(z)=\sum_{\nu }A_{\nu }\mathcal{E}_{i\nu }e^{+z/\ell _{\nu
}}+B_{\nu }\mathcal{E}_{i\nu }e^{-z/\ell _{\nu }}.  \label{SLs}
\end{equation}%
Substituting (\ref{SLs}) into BCs (\ref{BC.1}), we obtain
\begin{equation}
\sum_{\nu }\frac{\sigma _{\mathrm{D}}}{\ell _{\nu }}\left( A_{\nu }-B_{\nu
}\right) \mathcal{E}_{i\nu }=-\delta _{i2}2eJ_{\mathrm{SH}},
\end{equation}%
which leads into
\begin{equation}
\frac{A_{\nu }}{2eJ_{\mathrm{SH}}}-\frac{B_{\nu }}{2eJ_{\mathrm{SH}}}%
=-\mathcal{E}_{\nu 2}^{-1}\frac{\ell _{\nu }}{\sigma _{\mathrm{D}}}.
\end{equation}%
So that we obtain $\mu _{\mathrm{s}i}(z)=\mu _{\mathrm{s}i}^{\mathrm{ch}}(z)+\mu _{%
\mathrm{s}i}^{\mathrm{sH}}(z)$ with
\begin{equation}
\mu _{\mathrm{s}\imath }^{\mathrm{ch}}(z)=\sum_{\nu }A_{\nu }\mathcal{E}%
_{\imath \nu }2\cosh \left( \frac{z}{\ell _{\nu }}\right) ,  \label{Sch}
\end{equation}%
\begin{equation}
\mu _{\mathrm{s}\imath }^{\mathrm{sH}}(z)=2eJ_{\mathrm{SH}}\sum_{\nu }%
\mathcal{E}_{\imath \nu }\mathcal{E}_{\nu 2}^{-1}\frac{\ell _{\nu }}{\sigma
_{\mathrm{D}}}e^{-z/\ell _{\nu }}.  \label{Ssh}
\end{equation}%
Substituting the above solutions (\ref{Sch}) and (\ref{Ssh}) into BCs (\ref%
{BC.11}), we obtian
\begin{equation}
ej_{\mathrm{s}i}^{b}=\sum_{\nu }\mathcal{E}_{i\nu }A_{\nu } g_{\nu } 2\cosh \left( \frac{d_{N}}{\ell _{\nu }}\right).
\end{equation}%
with
\begin{equation}
g_{\nu }=\frac{\sigma _{\mathrm{D}}}{2\ell _{\nu }}\tanh \left( \frac{d_{N}}{%
\ell _{\nu }}\right) .
\end{equation}%
The current $j_{\mathrm{s}i}^{b}$ composes of two parts as follows
\begin{align}
    ej_{\mathrm{s}%
i}^{b}=-\delta _{i2}eJ_{\mathrm{SH}}+ej_{\mathrm{s}i}^{},
\end{align}
with
\begin{equation}
\frac{j_{\mathrm{s}i}^{}}{J_{\mathrm{SH}}}=\sum_{\nu }%
\mathcal{E}_{i\nu }\mathcal{E}_{\nu 2}^{-1} e^{-d_{N}/\ell _{\nu }}.
\end{equation}%
Thus, we obtain
\begin{equation}
A_{\nu } g_{\nu } 2\cosh \left( \frac{d_{N}}{%
\ell _{\nu }}\right) =\sum_{i}\mathcal{E}_{\nu i}^{-1}ej_{\mathrm{s}i}^{b},
\end{equation}%
which leads to
\begin{equation}
\frac{\mu _{\mathrm{s}i}^{\mathrm{ch}}(z)}{eJ_{\mathrm{SH}}}=\sum_{\nu }%
\mathcal{E}_{i\nu }\mathcal{E}_{\nu 2}^{-1}G_{\nu }^{-1}\frac{\cosh \left(
z/\ell _{\nu }\right) }{\cosh \left( d_{N}/\ell _{\nu }\right) },
\end{equation}%
with
\begin{equation}
G_{\nu }^{-1}=g_{\nu }^{-1}\left[
\mathrm{sech}\left( \frac{d_{N}}{\ell _{\nu }}\right) -1\right] -2\frac{\ell
_{\nu }}{\sigma _{\mathrm{D}}}e^{-d_{N}/\ell _{\nu }}.
\end{equation}%
Finally we obtain the total solutions
\begin{equation}
\frac{\mu _{\mathrm{s}i}\left( z\right) }{eJ_{\mathrm{SH}}}=\sum_{\nu }%
\mathcal{E}_{i\nu }\mathcal{E}_{\nu 2}^{-1}\left[ G_{\nu }^{-1}\frac{\cosh
\left( z/\ell _{\nu }\right) }{\cosh \left( d_{N}/\ell _{\nu }\right) }+2%
\frac{\ell _{\nu }}{\sigma _{\mathrm{D}}}e^{-z/\ell _{\nu }}\right] .
\label{PQ.1}
\end{equation}

Note that the charge density contains drift  and diffusive terms as follows 
\begin{align}
    \vec{j}_c=\left(1-2 \theta_{\text{SH}}^2\right) \sigma_D \vec{E}+\sum_a \theta_{\text{SH}} \hat{a} \times\left(-\mathcal{D} \boldsymbol{\nabla} \mu_s^a\right).
\end{align}
The total longitudinal (along $\hat{x}$) and transverse (along $\hat{%
y}$) charge currents are defined
\begin{align}
    J_L=\frac{1}{d_Z}\int^{d_Z}_{0} dz j^x_c (z,y),
\end{align}
\begin{align}
    J_T=\frac{1}{d_Z}\int^{d_Z}_{0} dz j^y_c (z,y).
\end{align}
Then, we obtain 
\begin{eqnarray} \label{PQ.3}
\frac{J_{L}}{J_{\mathrm{L0}}} &=&1-2 \theta_{\text{SH}}^2 -\theta _{\mathrm{SH}%
}^{2}\frac{\sigma _{\mathrm{D}}}{2d_{N}}\left[ \frac{\mu _{\mathrm{s}%
2}\left( d_{N}\right) }{ej_{\mathrm{sH}}^{0}}-\frac{\mu _{\mathrm{s}2}\left(
0\right) }{ej_{\mathrm{SH}}^{0}}\right]    \notag\\
&=&1-2 \theta_{\text{SH}}^2 -\theta _{\mathrm{SH}}^{2}\sum_{\nu }\mathcal{E}_{2\nu }\mathcal{E}_{\nu
2}^{-1}\zeta _{\nu },  
\end{eqnarray}%
\begin{eqnarray}
\frac{J_{T}}{J_{\mathrm{L0}}} &=&+\theta _{\mathrm{SH}}^{2}%
\frac{\sigma _{\mathrm{D}}}{2d_{N}}\left[ \frac{\mu _{\mathrm{s}1}\left(
d_{N}\right) }{ej_{\mathrm{sH}}^{0}}-\frac{\mu _{\mathrm{s}1}\left( 0\right)
}{ej_{\mathrm{sH}}^{0}}\right]   \label{PQ.4} \\
&=&+\theta _{\mathrm{SH}}^{2}\sum_{\nu }\mathcal{E}_{1\nu }\mathcal{E}_{\nu
2}^{-1}\zeta _{\nu },  \notag
\end{eqnarray}%
with $J_{\mathrm{L0}}=\sigma_DE$ where
\begin{equation}
\zeta _{\nu }=-\frac{2\ell _{\nu }}{d_{N}}\tanh \left( \frac{%
d_{N}}{2\ell _{\nu }}\right) ,
\end{equation}%
where $\zeta _{\nu }$ is the dimensionless quantities only depended on
mixing conductances. By means of  eigenvectors (\ref{EV}) and (\ref{IEV}), we obtain
\begin{align} \label{mmm1}
    \mathcal{E}_{2,1 }\mathcal{E}_{1,
2}^{-1}=\frac{+i\hat{m}_{3}+\hat{m}_{1}\hat{m}_{2}}{\sqrt{2}N_{r}}\frac{-i\hat{m}_{3}+\hat{m}_{1}%
\hat{m}_{2}}{\sqrt{2}N_{r}},
\end{align}
\begin{align}
    \mathcal{E}_{2,2}\mathcal{E}_{2,
2}^{-1}=\frac{-i\hat{m}_{3}+\hat{m}_{1}\hat{m}_{2}}{\sqrt{2}N_{r}}\frac{+i\hat{m}_{3}+\hat{m}_{1}%
\hat{m}_{2}}{\sqrt{2}N_{r}},
\end{align}
\begin{align}
    \mathcal{E}_{2,3 }\mathcal{E}_{3,
2}^{-1}=\hat{m}^2_2,
\end{align}
\begin{align}
    \mathcal{E}_{1,1 }\mathcal{E}_{1,
2}^{-1}=\frac{\hat{m}_{1}^{2}-1}{\sqrt{2}%
N_{r}}\frac{-i\hat{m}_{3}+\hat{m}_{1}%
\hat{m}_{2}}{\sqrt{2}N_{r}},
\end{align}
\begin{align}
    \mathcal{E}_{1,2}\mathcal{E}_{2,
2}^{-1}=\frac{\hat{m}_{1}^{2}-1}{\sqrt{2}%
N_{r}}\frac{+i\hat{m}_{3}+\hat{m}_{1}%
\hat{m}_{2}}{\sqrt{2}N_{r}},
\end{align}
\begin{align} \label{mmm6}
    \mathcal{E}_{1,3 }\mathcal{E}_{3,
2}^{-1}=\hat{m}_1\hat{m}_2.
\end{align}
Substituting Eqs. (\ref{mmm1}-\ref{mmm6})
into (\ref{PQ.3}) and (\ref{PQ.4}), we obtain
\begin{equation}
\frac{J_{L}}{J_{\mathrm{L0}}}=1-2 \theta_{\text{SH}}^2-\theta _{\mathrm{SH}%
}^{2}\left\{ \zeta _{3}+\left[ \mathrm{Re}\left( \zeta _{1}\right) -\zeta
_{3}\right] \left( 1-\hat{m}_{1}^{2}\right) \right\} ,
\end{equation}%
\begin{equation}
\frac{J_{T}}{J_{\mathrm{L0}}}=\theta _{\mathrm{SH}%
}^{2}\left\{ \left[ \zeta _{3}-\mathrm{Re}\left( \zeta _{1}\right) \right]
\hat{m}_{1}\hat{m}_{2}-\mathrm{Im}\left( \zeta _{1}\right) \hat{m}%
_{3}\right\} ,
\end{equation}%
which lead to
\begin{equation}
\frac{\rho _{L}}{\rho _{\mathrm{L0}}}\simeq \theta _{%
\mathrm{SH}}^{2}(2+\zeta _{3})+\theta _{\mathrm{SH}}^{2}\left[ \mathrm{Re}\left(
\zeta _{1}\right) -\zeta _{3}\right] \left( 1-\hat{m}_{2}^{2}\right) ,
\end{equation}%
\begin{equation}
\frac{\rho _{T}}{\rho _{\mathrm{L0}}}=-\theta _{\mathrm{SH}%
}^{2}\left[ \zeta _{3}-\mathrm{Re}\left( \zeta _{1}\right) \right] \hat{m}%
_{1}\hat{m}_{2}+\theta _{\mathrm{SH}}^{2}\mathrm{Im}\left( \zeta _{1}\right)
\hat{m}_{3}.
\end{equation}

\begin{figure}[t!]
\begin{center}
\includegraphics[width=0.48\textwidth]{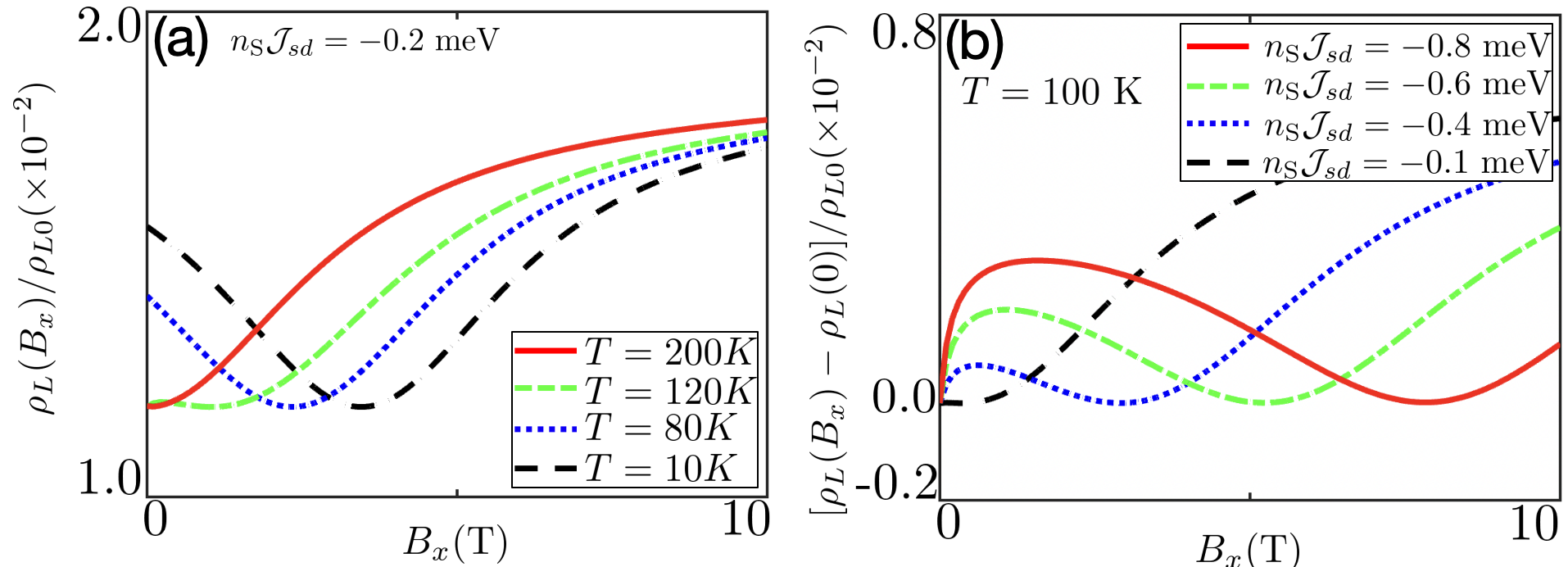} 
\end{center}
\caption{(Color online) Diverse MR behaviors. (a,b) Resistivity as a function of $\hat{x}$-axis
magnetic field, $B_{x}$, for different values of (a) temperature $T$  and (b) spin-exchange coupling $\mathcal{J}_{sd}$. (c,d) $\rho_{\text{L}}$ vs (c) $B_{x}$  and (d) $T$, for various $\mathcal{J}_{sd}$.  The $B$- and $T$-dependent $\mathcal{B}_{sd}$ causes a minimum in resistivity with $B$ (a-c) and $T$ (d) at $B_x=\mathcal{B}_{sd}$. Other parameters: $\theta _{%
\mathrm{SH}}=0.1$, $S=2$, $\ell _{0}=3.0$ nm, $d_{N}=5$ nm, $E_F=1.0 $ eV, $m_{F}=1.0$ $m^0_e$, $\rho_{L0}=2.0\times 10^{6}$ $\Omega\cdot m$, and $\mathcal{D}=1.0*10^{-6}$ m$^2$/s.}
\label{3DVACNJvsH}
\end{figure}

\section{Normal metal decorated with magnetic impurities} \label{abgvkfk}

In this Appendix, we present the MR effect of a normal metal decorated with magnetic impurities that are randomly distributed but still maintain  nonlocal spin-spin correlation to generate ferromagnetic configuration.  To quantitatively analyze the $B$- and $T$-dependent MR effect in ferromagnetic materials, our theory is numerically explored by considering a typical example - the Pt film decorated with magnetic permalloy (Py). Here, the spin-orbit coupling in Pt is characterized by a spin Hall angle $\theta_{\mathrm{SH}}\simeq 0.1$ \cite{ando2008electic} and spin diffusion length $\ell _{0}=3.0n$m comparable to the thickness of Pt film, $d_N = 5$ nm. The local moments (Py) have spin $S=2$ and density $n_{\mathrm{S}}a^3_{\text{Pt}}=0.1$ where  $a_{\text{Pt}}$ is the lattice constant of Pt and $n_{\mathrm{S}}$ should be high enough to generate ferromagnetic order of local moments.  Moreover, our spin-exchange coupling $n_{\text{S}}\mathcal{J}_{sd}$ is set to be the order of meV~\footnote{The spin-exchange coupling between Co adatoms on Cu(100) has been predicted to realize a ferromagnetic interaction of about $n_{\text{S}}\mathcal{J}_{sd}\simeq 350$ meV and an antiferromagnetic one about $n_{\text{S}}\mathcal{J}_{sd}\simeq -17 $ meV \citep{stepanyuk2001magnetic,wahl2007exchange}, which have been validated by probing the Kondo resonance in experiments of P. Wahl $et. al$ \cite{wahl2007exchange}. Though there is no experimental data of spin-exchange coupling in Pt film decorated with Py, we believe $n_{\text{S}}\mathcal{J}_{sd}\sim$ meV should be an experimentally feasible parameter.}. Below, we discuss isotropic magnon, anisotropic spin precession and spin relaxation MR together with diverse MR behaviors according to our universal formulas (\ref{SL}-\ref{vvh1}).

Next, we show intriguing MR for the magnetic field in $x$-axis direction ($B_x$). Figure~\ref{3DVACNJvsH}(a) plots a transition from positive to
negative MR at small $B$ for antiferromagnetic spin-exchange coupling when the system is cooled from a high $
(T=200K)$ to low  temperatures $(T=10K)$. This interesting transition can be explained as
follows.  The spin-exchange coupling \eqref{spin-exchange} induces 
a spin-exchange field \eqref{elf}, which is linearly proportional to the spin-exchange coupling $\mathcal{J}_{sd}$ and the magnetization $
\langle S_{\parallel }\rangle$, i.e., $\mathcal{B}_{sd}\propto \mathcal{J}_{sd}\langle S_{\Vert }\rangle$. Hence, the ferromagnetic (antiferromagnetic) spin-exchange coupling generates a
blue (red) shift of the Hanle spin precession frequency $\omega _{L}=\textsl{g}\mu _{B}(B_x-\mathcal{B}_{sd})/\hbar$ with $\mathcal{B}_{sd}<0$ ($\mathcal{B}_{sd}>0$). For high enough temperature (red curve), $\mathcal{B}_{sd}$ vanispin Hall effects because $\langle S_{\parallel }\rangle\simeq 0$,  and we recover the previous Hanle MR that exhibits quadratic behavior concerning small $B_x$, i.e., $\rho_{\text{L}}\propto \omega^2_L\propto B_x^2$ \cite{velez2016hanle}. Thus, we always get positive MR for ferromagnetic and antiferromagnetic spin-exchange coupling.  At a low enough temperature (black curve), i.e., strongly magnetized regime, the spin-exchange field \eqref{elf}, spin relaxation time \eqref{LongiSRT} and \eqref{TransSRT} acquire their saturated values and become independent of the magnetic field. Then, the magnetic field dependence of resistance purely originates from the magnetic-field spin precession frequency, leading to the shifted spin precession MR $\rho_{\text{L}}\propto \omega^2 _{L}\propto  (B_x-\mathcal{B}_{sd})^2$ with $
\mathcal{B}_{sd} = -n_{\mathrm{S}}   \mathcal{J}_{sd}   S/(\hbar g\mu _{B})$. The ferromagnetic (antiferromagnetic) spin-exchange coupling leads to positive (negative) MR with $\mathcal{B}_{sd}<0$ $(\mathcal{B}_{sd}>0)$, as indicated by the red and blue curves in Fig.~\ref{3DVACNJvsH}(b). Thus, we find a transition from positive to negative MR for antiferromagnetic spin-exchange coupling by cooling
the system to low temperatures.

\end{document}